\documentclass[aps,showpacs,pre,twocolumn]{revtex4-1}

\usepackage{graphicx}
\usepackage{amsmath}
\pdfoutput=1

\begin{document}

\title{Collective motion of binary self-propelled particle mixtures}

\date{\today}

\author{Andreas M.~Menzel}
\altaffiliation[Current address: ]{Institut f\"ur Theoretische Physik II: Weiche Materie, Heinrich-Heine-Universit\"at, Universit\"atsstra{\ss}e 1, D-40225 D\"usseldorf, Germany}
\email[email: ]{menzel@thphy.uni-duesseldorf.de}
\affiliation{Max Planck Institute for Polymer Research, P.O.~Box 3148, 55021 Mainz, Germany}

\begin{abstract}
In this study, we investigate the phenomenon of collective motion in binary mixtures of self-propelled particles. More precisely, we consider two particle species, each of which consisting of pointlike objects that propel with a velocity of constant magnitude. Within each species, the particles try to achieve polar alignment of their velocity vectors, whereas we analyze the cases of preferred polar, antiparallel, as well as perpendicular alignment between particles of different species. Our focus is on the effect that the interplay between the two species has on the threshold densities for the onset of collective motion and on the nature of the solutions above onset. For this purpose, we start from suitable Langevin equations in the particle picture, from which we derive mean field equations of the Fokker-Planck type and finally macroscopic continuum field equations. We perform particle simulations of the Langevin equations, linear stability analyses of the Fokker-Planck and macroscopic continuum equations, and we numerically solve the Fokker-Planck equations. Both, spatially homogeneous and inhomogeneous solutions are investigated, where the latter correspond to stripe-like flocks of collectively moving particles. In general, the interaction between the two species reduces the threshold density for the onset of collective motion of each species. However, this interaction also reduces the spatial organization in the stripe-like flocks. The case that shows the most interesting behavior is the one of preferred perpendicular alignment between different species. There, a competition between polar and truly nematic orientational ordering of the velocity vectors takes place within each particle species. Finally, depending on the alignment rule for particles of different species and within certain ranges of particle densities, identical and inverted spatial density profiles can be found for the two particle species. The system under investigation is confined to two spatial dimensions. 
\end{abstract}



\maketitle

\section{Introduction}

Inspired by the biological observation of microorganisms, self-propulsion has widely been studied as the motion of microswimmers through their viscous fluid environment. Different models were suggested and analyzed for this kind of self-propulsion in the low Reynolds number limit. Examples are active rods propagating planar or spiral waves along their body \cite{taylor1951analysis,hancock1953self}, Purcell's swimmer that consists of three bars connected by two joints \cite{purcell1977life,becker2003self}, a model swimmer of three linked spheres \cite{najafi2004simple}, or helical filaments that propagate kinks between regions of opposite handedness through their body \cite{wada2007model}. 

Apart from the propulsion mechanism of single isolated objects, their collective behavior under hydrodynamic coupling has been moved into the focus of recent investigations. For example, hydrodynamic continuum equations have been derived by coarse-graining the particle picture of interacting microswimmers \cite{baskaran2009statistical}. In this case, orientational order parameters follow from locally averaging the single particle velocities. The mode structure and hydrodynamic instabilities of the corresponding continuum equations were analyzed \cite{aranson2007model,baskaran2009statistical,leoni2010swimmers}. Here, we can identify the thresholds at which orientationally non-ordered states become unstable with the onset of collective motion. Aspects of synchronization of active particle motion mediated by hydrodynamic interactions have been studied in detail \cite{uchida2010synchronization,uchida2011generic,uchida2011many,golestanian2011hydrodynamic}. Furthermore, a recent work incorporates the effect of an axial macroscopic dynamic variable into the hydrodynamic framework \cite{brand2011macroscopic}. Such a macroscopic non-equilibrium variable may appear, for example, in the case of collectively rotating helical filaments. 

In this paper, we will restrict ourselves to collective two-dimensional motion of self-propelled particles on a substrate. Therefore, no long-range hydrodynamic interactions are taken into account. Such a system was used, for example, to study the motion of deformable self-propelled particles and domains \cite{ohta2009deformable,ohkuma2010deformable,hiraiwa2010dynamics,itino2011collective}. There, the velocity of each particle was determined through coupling to its deformation (and vice versa). In contrary, in simpler models, a constant driving force is directly added to the momentum equation for each particle \cite{levine2000self,peruani2006nonequilibrium,baskaran2008enhanced}. It balances the friction forces acting on the particle. The driving force is oriented, for example, along the long axis of rod-like particles \cite{peruani2006nonequilibrium,baskaran2008enhanced}. Shape dependent short-range interactions as, e.g., excluded volume interactions can lead to alignment and thus collective motion. Again, macroscopic equations were derived and investigated for these cases \cite{baskaran2008enhanced}. 

When we are only interested in the basic mechanism that leads to collective motion, we only keep the basic necessary features in a minimal model. Then the velocity magnitude for each particle can simply be kept fixed as a constant in time \cite{baskaran2008hydrodynamics}. Only the position and the angle characterizing its direction of motion are maintained as variables for each particle. For such a system, we can define local alignment rules for the velocity vectors of two interacting particles \cite{gregoire2004onset,chate2006simple,bertin2006boltzmann,bertin2009hydrodynamic, ginelli2010relevance,ginelli2010large}. Since momentum exchange with the substrate is possible, the total momentum of the interacting particles does not need to be conserved. 

Such approaches can be seen as variants of the model introduced by Vicsek et al. \cite{vicsek1995novel,chate2008modeling} and were extensively studied numerically. Both, polar \cite{gregoire2004onset,bertin2006boltzmann,bertin2009hydrodynamic,chate2008collective, ginelli2010relevance,peruani2010cluster,peruani2011polar} and nematic \cite{chate2006simple,ginelli2010large,peruani2010cluster,peruani2011polar} alignment rules were investigated. In the first case, velocity vectors in the ordered state point into the same direction, whereas in the second case antiparallel alignment of velocity vectors is equally allowed. The focus was on the transition from disordered to collective, i.e.~macroscopically ordered, motion. This transition occurs with decreasing orientational noise or increasing particle density and was found to be discontinuous for sufficiently large system sizes \cite{gregoire2004onset,chate2008collective}. An interesting issue was the emergence of spatial heterogeneities above (but close to) threshold \cite{gregoire2004onset,chate2006simple,chate2008collective,chate2008modeling,peruani2010cluster, ginelli2010large,peruani2011polar}. More precisely, flocks and stripes of high particle density and orientational order were observed. They could move with velocity magnitudes close to the single particle speed. 

Apart from that, hydrodynamic continuum equations were derived from alignment rules through a Boltzmann approach \cite{bertin2006boltzmann,bertin2009hydrodynamic}. Furthermore, continuum equations of the Fokker-Planck type were obtained using the mean field approximation \cite{peruani2008mean,lee2010fluctuation}. The features listed above could be reproduced in numerical investigations of macroscopic continuum equations \cite{mishra2010fluctuations,toner2005hydrodynamics}. 

We will follow in this paper the outlined minimal model approach. The system that we will study is a binary mixture of self-propelled particles. Previously, systems of different self-propelled particles were analyzed in predator-prey scenarios \cite{mecholsky2010obstacle,sengupta2011chemotactic}. In the current work, however, we focus on the effect of alignment interactions between different particle species. More precisely, we concentrate on the onset and features of collective motion that result from the interaction between two different groups of particles. Therefore, in our analysis, the main parameter of interest will be the orientational coupling parameter between particles of different species. 

In the next section, we introduce in detail our minimal model of binary self-propelled particle mixtures on the particle level. From these, we derive mean field equations in the spirit of the Fokker-Planck approach. After that, in section \ref{linstabanal}, these equations are checked for linear instabilities of orientational order as a function of the averaged particle densities. We study the cases of polar and antiparallel alignment of the velocity vectors as well as a preferred perpendicular alignment. These cases were further investigated by particle simulations, the results of which are presented in section \ref{particleperp}. After that, we compare to numerical results obtained from the Fokker-Planck approach in section \ref{numFP}. Finally, we derive macroscopic continuum equations in section \ref{macroscopic} and discuss the stability of their solutions, before we conclude.

\section{Particle and field description}\label{particlefield}

In the following, we describe our two-dimensional minimal model of a binary mixture of self-propelled particles. We consider $N$ interacting particles. From these $N$ particles, $N_1$ particles are of species $1$, the other $N-N_1$ particles are of species $2$. Each particle is assumed to propel with a velocity of fixed constant magnitude. More precisely, particles of species $1$ and $2$ propel with a velocity of constant magnitude $2u_1$ and $2u_2$, respectively. The orientation of the velocity vector $\mathbf{v}_i$ of each particle $i$ ($i=1,2,\dots,N$) can be characterized by a single orientation angle $\theta_i$ in the two-dimensional plane. We measure these angles $\theta_i$ with respect to the positive $x$-axis. The independent variables in our model are thus the position vectors $\mathbf{r}_i$ and the velocity orientation angles $\theta_i$ of all particles, $i=1,2,\dots,N$. 

These variables follow a set of coupled Langevin equations
\begin{eqnarray}
\frac{d\mathbf{r}_i}{dt} &=& \mathbf{v}_i(\theta_i) = 2u_{(i)}
  \left(\begin{array}{c} \cos\theta_i \\ \sin\theta_i \end{array}\right), \label{updater}\\
\frac{d\theta_i}{dt} &=& {}-\frac{\partial U(\mathbf{r}_1,\dots,\mathbf{r}_N,\theta_1,\dots,\theta_N)}{\partial\theta_i} +\Gamma_i(t). \label{updatetheta}
\end{eqnarray}
Here, $u_{(i)}=u_1$ for $i=1,\dots,N_1$ and $u_{(i)}=u_2$ for $i=N_1+1,\dots,N$. $\Gamma_i(t)$ is a Gaussian stochastic force of zero mean and correlation of the form $\langle\Gamma_i(t)\Gamma_j(t')\rangle=2D_i\delta_{ij}\delta(t-t')$. Likewise, we have $D_i=D_1$ for $i=1,\dots,N_1$ and $D_i=D_2$ for $i=N_1+1,\dots,N$. 

In this context, $U(\mathbf{r}_1,\dots,\mathbf{r}_N,\theta_1,\dots,\theta_N)$ plays the role of an interaction potential between the particles. It controls the preferred orientational alignment (or misalignment) between particles of identical and different species. We assume the following form of a pairwise particle-particle interaction potential,
\begin{eqnarray}\label{U}
\lefteqn{U(\mathbf{r}_1,\dots,\mathbf{r}_N,\theta_1,\dots,\theta_N)} \nonumber\\[.2cm]
&=& {}-\frac{g_1}{\pi}\sum_{i=1}^{N_1-1}\sum_{j=i+1}^{N_1}\delta(\mathbf{r}_i-\mathbf{r}_j)\cos(\theta_i-\theta_j) \nonumber\\
&& {}-\frac{g_2}{\pi}\sum_{i=N_1+1}^{N-1}\sum_{j=i+1}^{N}\delta(\mathbf{r}_i-\mathbf{r}_j)\cos(\theta_i-\theta_j) \nonumber\\
&& {}-\frac{g}{\pi}\sum_{i=1}^{N_1}\sum_{j=N_1+1}^{N}\delta(\mathbf{r}_i-\mathbf{r}_j)U_g(\theta_i-\theta_j). 
\end{eqnarray}
Here, we assume that $g_1>0$ and $g_2>0$. The terms with the coefficients $g_1$ and $g_2$ make two particles of the same species try to align their velocity vectors parallel to each other and pointing into the same direction (polar alignment). In this study, we restrict ourselves to idealized local point-like interactions, which leads to the spatial $\delta$-functions in Eq.~(\ref{U}). 

\begin{figure}
\includegraphics[width=8.5cm]{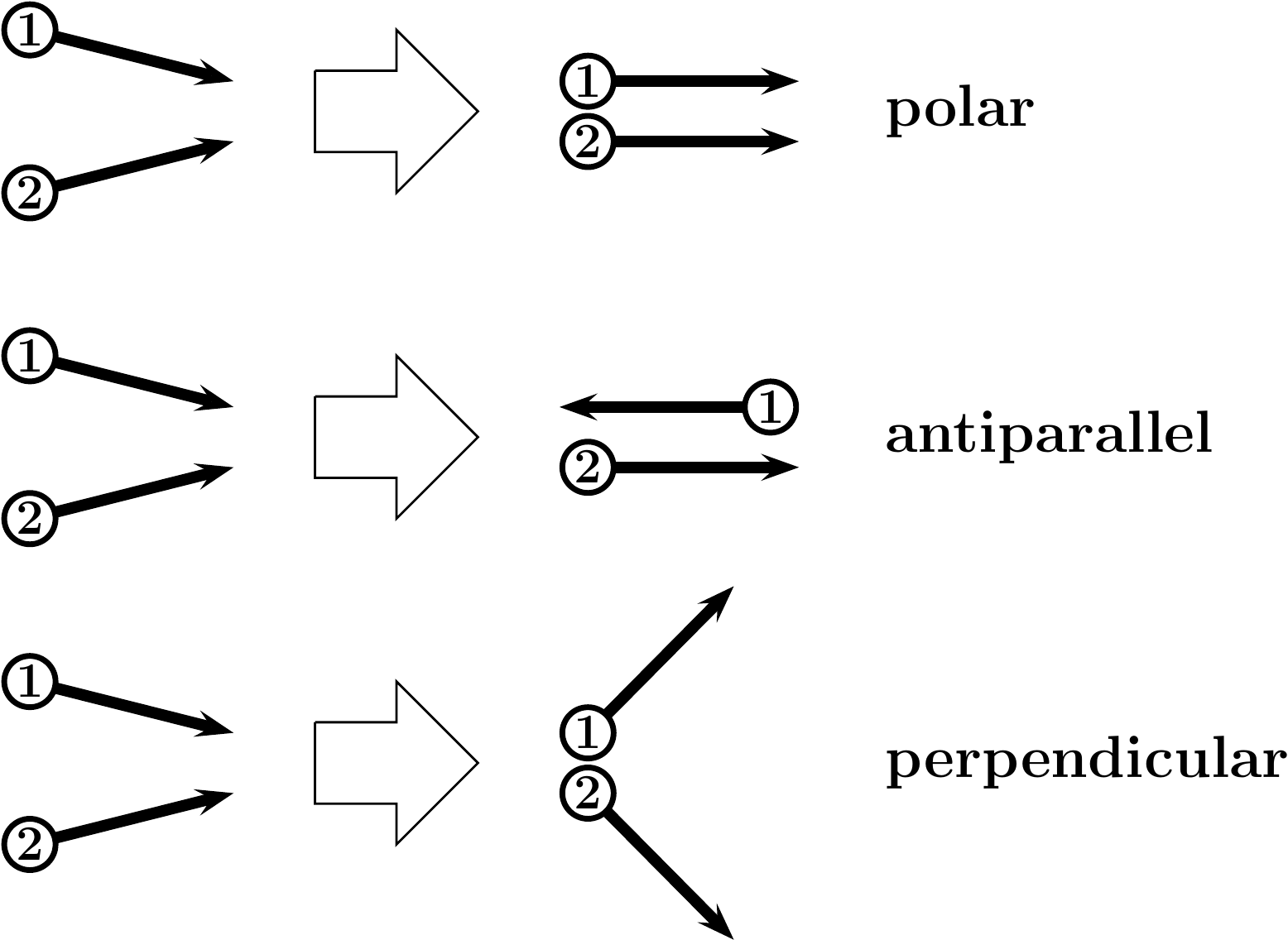}
\caption{Illustration of polar, antiparallel, and perpendicular alignment interactions between particles of the different species $1$ and $2$.} 
\label{alignmentrules}
\end{figure}

For particles of different species, we consider the three different alignment rules depicted in Fig.~\ref{alignmentrules}. 
Parallel alignment of the velocity vectors is achieved through the angular interaction potential 
\begin{equation}\label{Ugpar}
U_g(\theta_i-\theta_j)=U_g^{\|}(\theta_i-\theta_j)=\cos(\theta_i-\theta_j). 
\end{equation}
We find preferred polar alignment for $g>0$ and preferred antiparallel alignment for $g<0$. Due to the idealized local point-like interactions, antiparallel alignment does not lead to spatial repulsion. Flocks of different particle species can simply penetrate each other. In this case, for the two species to spatially avoid each other, spatially nonlocal repulsive forces and/or excluded volume forces must be taken into account. 

In the idealized picture, for particles of different species to try to spatially avoid each other, the preferred alignment of their velocity vectors must not be parallel to each other. We investigate preferred perpendicular alignment, which is supported by the term with the coefficient $g$ via 
\begin{equation}\label{Ugperp}
U_g(\theta_i-\theta_j)=U_g^{\perp}(\theta_i-\theta_j)=\sin^2(\theta_i-\theta_j)
\end{equation}
and $g>0$.

We now ask for a temporal evolution equation for the probability density $f(\mathbf{r}_1,\dots,\mathbf{r}_N,\theta_1,\dots,\theta_N,t)$ of finding simultaneously particles $1,\dots,N$ at positions $\mathbf{r}_1,\dots,\mathbf{r}_N$ and with velocity orientations $\theta_1,\dots,\theta_N$. Through the well-known procedures \cite{risken1996fokker,zwanzig2001nonequilibrium} we obtain an equation of the Fokker-Planck type that reads
\begin{eqnarray}
\lefteqn{\frac{\partial f(\mathbf{r}_1,\dots,\mathbf{r}_N,\theta_1,\dots,\theta_N,t)}{\partial t}}    \nonumber\\
&=& \sum_{i=1}^{N}\bigg\{ -2u_i
\big[\cos\theta_i\,\partial_x+\sin\theta_i\,\partial_y\big]    \nonumber\\
&& \hspace{.5cm}{}+\partial_{\theta_i}\frac{\partial U(\mathbf{r}_1,\dots,\mathbf{r}_N,\theta_1,\dots,\theta_N)}{\partial\theta_i}    \nonumber\\ 
&& \hspace{.5cm}{}+D_i\partial_{\theta_i}^2\bigg\}\,f(\mathbf{r}_1,\dots,\mathbf{r}_N,\theta_1,\dots,\theta_N,t).
\end{eqnarray}

From this equation, we find the time evolution of the one-particle density $\rho_1^{(1)}(\mathbf{r},\theta,t)$ by integrating out all variables $\mathbf{r}_2,\dots,\mathbf{r}_N,\theta_2,\dots,\theta_N$, identifying the remaining variables $\mathbf{r}_1\equiv\mathbf{r}$ and $\theta_1\equiv\theta$, and multiplying by $N_1$. Here, the superscript of $\rho_1^{(1)}$ denotes the one-particle density, whereas the subscript refers to species $1$. Likewise, we derive a dynamic equation for the one-particle density $\rho_2^{(1)}(\mathbf{r},\theta,t)$ of species $2$.

These equations contain the two-particle densities. For example, the two-particle density $\rho^{(2)}_{11}(\mathbf{r},\mathbf{r'},\theta,\theta')$ is obtained from $f(\mathbf{r}_1,\dots,\mathbf{r}_N,\theta_1,\dots,\theta_N,t)$ by integrating out all variables $\mathbf{r}_3,\dots,\mathbf{r}_N,\theta_3,\dots,\theta_N$, identifying the remaining variables $\mathbf{r}_1\equiv\mathbf{r}$, $\mathbf{r}_2\equiv\mathbf{r'}$, $\theta_1\equiv\theta$, as well as $\theta_2\equiv\theta'$, and multiplying by $N_1(N_1-1)$. To close the equations, we apply the mean field approximation for the two-particle densities 
$\rho^{(2)}_{11}(\mathbf{r},\mathbf{r'},\theta,\theta')=\rho^{(1)}_1(\mathbf{r},\theta)\rho^{(1)}_1(\mathbf{r'},\theta')$, 
$\rho^{(2)}_{12}(\mathbf{r},\mathbf{r'},\theta,\theta')=\rho^{(1)}_1(\mathbf{r},\theta)\rho^{(1)}_2(\mathbf{r'},\theta')$, 
$\rho^{(2)}_{21}(\mathbf{r},\mathbf{r'},\theta,\theta')=\rho^{(1)}_2(\mathbf{r},\theta)\rho^{(1)}_1(\mathbf{r'},\theta')$, 
and $\rho^{(2)}_{22}(\mathbf{r},\mathbf{r'},\theta,\theta')=\rho^{(1)}_2(\mathbf{r},\theta)\rho^{(1)}_2(\mathbf{r'},\theta')$. The analogous procedure was applied before in the case of a single particle species \cite{bertin2006boltzmann,bertin2009hydrodynamic,lee2010fluctuation}. In ref.~\cite{savel2003controlling}, a similar procedure is applied to the conventional case of non-propelled Brownian particle mixtures. 

Since we are only referring to one-particle densities, we omit the superscript $^{(1)}$ during the rest of this manuscript. The resulting equations read
\begin{eqnarray}\label{eq1}
\lefteqn{\frac{\partial\rho_1(\mathbf{r},\theta,t)}{\partial t}\,=} \nonumber\\[.1cm]
&& {}-2u_1\big[ \cos\theta\,\partial_x+\sin\theta\,\partial_y \big]\rho_1(\mathbf{r},\theta,t) \nonumber\\[.1cm]
&& {}+D_1\partial_{\theta}^2\rho_1(\mathbf{r},\theta,t) \nonumber\\[.1cm]
&& {}+\frac{g_1}{\pi}\partial_{\theta}\int_0^{2\pi}\sin(\theta-\theta')\rho_1(\mathbf{r},\theta,t)\rho_1(\mathbf{r},\theta',t)d\theta' \nonumber\\
&& {}+\frac{g}{\pi}\partial_{\theta}\int_0^{2\pi}\sin[a(\theta-\theta')]\rho_1(\mathbf{r},\theta,t)\rho_2(\mathbf{r},\theta',t)d\theta', \hspace{.6cm}
\end{eqnarray}
\begin{eqnarray}\label{eq2}
\lefteqn{\frac{\partial\rho_2(\mathbf{r},\theta,t)}{\partial t}\,=} \nonumber\\[.1cm]
&& {}-2u_2\big[ \cos\theta\,\partial_x+\sin\theta\,\partial_y \big]\rho_2(\mathbf{r},\theta,t) \nonumber\\[.1cm]
&& {}+D_2\partial_{\theta}^2\rho_2(\mathbf{r},\theta,t) \nonumber\\[.1cm]
&& {}+\frac{g_2}{\pi}\partial_{\theta}\int_0^{2\pi}\sin(\theta-\theta')\rho_2(\mathbf{r},\theta,t)\rho_2(\mathbf{r},\theta',t)d\theta' \nonumber\\
&& {}+\frac{g}{\pi}\partial_{\theta}\int_0^{2\pi}\sin[a(\theta-\theta')]\rho_2(\mathbf{r},\theta,t)\rho_1(\mathbf{r},\theta',t)d\theta'. \hspace{.6cm}
\end{eqnarray}
Here, $a=-2$ for Eq.~(\ref{Ugperp}) and $a=1$ in case of Eq.~(\ref{Ugpar}).

It can be seen from Eqs.~(\ref{updater})--(\ref{U}), (\ref{eq1}), and (\ref{eq2}) that $u_j$ ($j=1,2$) have dimensions of velocity and $D_j$ ($j=1,2$) have dimensions of inverse time. From dimensional analysis it follows that the parameters $g_1$, $g_2$, and $g$ can be measured in units of $u_1^2/D_1$ or $u_2^2/D_2$. 

Later in this first study, we will restrict ourselves to the case where the particles of the two species feature identical physical behavior. More precisely, this means $u_1=u_2$, $D_1=D_2$, and $g_1=g_2$. In this case, we can set $u_1=u_2=1$, $D_1=D_2=1$, and $g_1=g_2=1$ without loss of generality. The latter can be seen, for example, from rescaling time, space, and densities via $t\rightarrow t'/D_1$, $\mathbf{r}\rightarrow\mathbf{r'}u_1/D_1$, $\rho_1\rightarrow\rho_1'D_1/g_1$, and $\rho_2\rightarrow\rho_2'D_1/g_2$. As a consequence, there are only four different parameters remaining, namely $u_2/u_1$, $D_2/D_1$, $g/g_1$, and $g/g_2$. For the special case of identical physical behavior of the two particles there remains only one independent parameter that we can define as $g'=g/g_1=g/g_2$.

\section{Linear stability analysis}\label{linstabanal}

We see from Eqs.~(\ref{eq1}) and (\ref{eq2}) that $\rho_1(\mathbf{r},\theta,t)\equiv\rho_{10}$ and $\rho_2(\mathbf{r},\theta,t)\equiv\rho_{20}$ is always a solution. Here, $\rho_{10}$ and $\rho_{20}$ give the spatially and angularly averaged densities for the two species. They are conserved quantities. 

As a first step, we check the linear stability of this trivial solution with respect to orientational order. This is interesting because the points of instability of the trivial solution correspond to the onset of collective motion. 

We will not explicitly include the spatial component in these considerations. The gradient term only leads to an additional imaginary contribution to the listed eigenvalues. Therefore it does not modify the location of the obtained threshold points at which collective motion sets in.  

\subsection{Polar and antiparallel alignment}

To check linear stability for the case of polar and antiparallel alignment rules between different particle species, we must choose $a=1$ corresponding to Eq.~(\ref{Ugpar}). We insert the ansatz
\begin{equation}\label{ansatz-spat-hom}
\rho_j(\mathbf{r},\theta,t)=\rho_{j0}+\tilde{\rho}_{j0}e^{in\theta+\lambda t}, \quad j=1,2
\end{equation}
into Eqs.~(\ref{eq1}) and (\ref{eq2}). Linearizing in the amplitudes $\tilde{\rho}_{j0}$, $j=1,2$, leads to
\begin{eqnarray}
\lambda\tilde{\rho}_{10} &=& {}-D_1n^2\tilde{\rho}_{10} +(g_1\tilde{\rho}_{10} +g\tilde{\rho}_{20})\rho_{10}\delta_{n1}, \label{antiparlambda1}\\
\lambda\tilde{\rho}_{20} &=& {}-D_2n^2\tilde{\rho}_{20} +(g_2\tilde{\rho}_{20} +g\tilde{\rho}_{10})\rho_{20}\delta_{n1}. \label{antiparlambda2}
\end{eqnarray}
Therefore, only the mode $n=1$ can become linearly unstable. 

From these equations, we obtain the eigenvalues
\begin{eqnarray}\label{eigenvalpar}
\lefteqn{\lambda = \frac{1}{2}\bigg\{(g_1\rho_{10}-D_1)+(g_2\rho_{20}-D_2)} \nonumber\\[.1cm]
&& {}\pm\sqrt{\left[(g_1\rho_{10}-D_1)-(g_2\rho_{20}-D_2)\right]^2+4g^2\rho_{10}\rho_{20}}\bigg\}.
\nonumber\\
&& 
\end{eqnarray}
On increasing the average densities, the ``$+$''-eigenvalue always becomes unstable first and defines the onset of the linear instability. Interestingly, the sign of $g$ does not play a role. Therefore, the two vectors of collective motion of the two different species can align in a polar or antiparallel way at onset. Without any coupling ($g=0$), we correctly recover the behavior of the separate single-component systems $1$ and $2$ with the respective eigenvalues $\lambda_j=g_j\rho_{j0}-D_j$, $j=1,2$. This case leads to the well-known threshold densities $\rho_{j0}^{*,1}$ for the single-component systems: we find collective motions for densities higher than $\rho_{j0}^{*,1}=D_j/g_j$, $j=1,2$. 

Clearly, if both $\rho_{10}$ and $\rho_{20}$ are higher than the threshold single-component densities, we also find collective motion in the two-component system. Can collective motion also set in, however, if both $\rho_{10}$ and $\rho_{20}$ are smaller than the threshold single-component densities? 

Analysis of Eq.~(\ref{eigenvalpar}) shows that for this purpose
\begin{equation}\label{thresholdrho10}
\rho_{10}\left[(g_1g_2-g^2)\rho_{20}-D_2g_1\right]<D_1(g_2\rho_{20}-D_2)
\end{equation}
is a necessary condition. On the one hand, if $g^2>g_1g_2$, $\rho_{20}$ can have any value, and collective motion sets in at a lower density $\rho_{10}$ than in the single-component case. The threshold density $\rho_{10}^{*,2}$ is then given by condition (\ref{thresholdrho10}). On the other hand, if $g^2<g_1g_2$, it follows from Eq.~(\ref{eigenvalpar}) that both $\rho_{10}$ and $\rho_{20}$ can be smaller than their threshold single-component values and induce collective motion. In detail, this can be seen by rewriting Eq.~(\ref{thresholdrho10}) for the case that $\rho_{20}<D_2g_1/(g_1g_2-g^2)$ to
\begin{equation}\label{thresholdrho10rewritten}
\rho_{10}>\frac{D_1}{g_1}\frac{g_2\rho_{20}-D_2}{(g_2-g^2/g_1)\rho_{20}-D_2}. 
\end{equation}
For $\rho_{20}=0$ we recover the single-component condition for $\rho_{10}$. When we now increase $\rho_{20}$, the (negative) numerator in Eq.~(\ref{thresholdrho10rewritten}) grows faster than the (negative) denominator because $g_2>g_2-g^2/g_1>0$. 

As a result, the two species support each other in starting to move collectively. In this way, the critical threshold densities can be smaller than for the single-component systems. Of course, the above analysis analogously applies when the two species are switched by changing the subscripts $1\leftrightarrow 2$.

\subsection{Perpendicular alignment}

Considering now $a=-2$ and again ansatz (\ref{ansatz-spat-hom}) in Eqs.~(\ref{eq1}) and (\ref{eq2}), we obtain 
\begin{eqnarray}
\lambda\tilde{\rho}_{10} &=& {}-D_1n^2\tilde{\rho}_{10} +g_1\rho_{10}\tilde{\rho}_{10}\delta_{n1} \nonumber\\
&& {}-2g\rho_{10}\tilde{\rho}_{20}\delta_{n2}, \label{perplambda1}\\
\lambda\tilde{\rho}_{20} &=& {}-D_2n^2\tilde{\rho}_{20} +g_2\rho_{20}\tilde{\rho}_{20}\delta_{n1} \nonumber\\
&& {}-2g\rho_{20}\tilde{\rho}_{10}\delta_{n2}. \label{perplambda2}
\end{eqnarray}
Consequently, now the modes $n=1$ and $n=2$ can become unstable. 

On the one hand, mode $n=1$ becomes unstable if at least one of the two averaged densities is higher than the threshold one 
\begin{equation}\label{critrhosep}
\rho_{j0}^{*,1} = \frac{D_j}{g_j}, \quad j=1,2. 
\end{equation}
These are the same values as for each one-species system separately. 

On the other hand, mode $n=2$ becomes unstable for the density product $\rho_{10}\rho_{20}$ larger than the threshold product
\begin{equation}\label{mod2instab}
\rho_{10}^{*,2} \rho_{20}^{*,2} = 4\frac{D_1D_2}{g^2}.
\end{equation}
To become unstable before the mode $n=1$ and thus below the density threshold of the single-component system, we must have $\rho_{j0}^{*,2}<\rho_{j0}^{*,1}$, $j=1,2$. For this to be possible, the parameters must satisfy the condition
\begin{equation}\label{gthresh}
g^2>4g_1g_2.
\end{equation}

It follows from Eqs.~(\ref{perplambda1}) and (\ref{perplambda2}) that the instability of the mode $n=2$ leads to $\tilde{\rho}_{10}$ and $\tilde{\rho}_{20}$ having opposite signs. Consequently, particles of species $1$ then on average propel perpendicularly to particles of species $2$. We can say that different species try to evade each other by perpendicular alignment of their velocity vectors. It is interesting to note that the addition of a strongly avoiding species can reduce the critical density at which collective motion sets in.

\section{Particle simulations}\label{particleperp}

As we have seen at the end of the last section, the case of perpendicular alignment of the velocity vectors of the different species is more complex than the case of polar and antiparallel alignment. Two alignment modes compete at onset, corresponding to polar (mode $n=1$) and nematic (mode $n=2$) orientational order. We can identify the coupling parameter $g$ as the crucial parameter that controls the nature of the orientational mode. To get more insight into the corresponding relations, we have performed particle simulations of binary self-propelled particle mixtures. Here, we mainly report our results as a function of the coupling parameter $g$. According to the rescaling procedure pointed out at the end of section \ref{particlefield}, the inter-species coupling parameter $g$ will be given in units of the intra-species coupling parameters $g_1=g_2$. 

We chose a two-dimensional quadratic simulation area of size $L_x\times L_y$ with periodic boundary conditions and $N_j$ particles of species $j$ ($j=1,2$). The quadratic shape was used to offer symmetric directions in the case of global perpendicular alignment. For simplicity, the velocities were set to $u_1=u_2=1$ and the interaction parameters between particles of the same species to $g_1=g_2=1$. During each iterating time step, the orientation angle $\theta_i$ of each particle was updated according to Eq.~(\ref{updatetheta}). In contrast to Eq.~(\ref{U}), however, particle interactions were not completely localized for practical reasons. We chose a disk-like environment of radius $d_0$ around each particle, where we mostly used $d_0=0.05$. Interactions were considered between particles within this distance. For this reason, results from the particle simulations cannot be quantitatively compared with the results from the Fokker-Planck approach. Apart from that, at each time step, we disturbed the orientation angle of each particle by an additive Gaussian stochastic noise term according to Eq.~(\ref{updatetheta}). The variance of the Gaussian distribution was chosen such that the angular diffusion parameters were $D_1=D_2=1$. After that, the advection step was performed according to Eq.~(\ref{updater}). The time increment was set to $dt=0.1$. Following our rescaling as mentioned at the end of section \ref{particlefield}, distances such as $d_0$ are measured in units of $u_1/D_1=u_2/D_2$ and time steps in units of $1/D_1=1/D_2$. 

As an initial condition, a random spatial and angular distribution of the particles was chosen. However, spatial overlap of the surrounding disks of radius $d_0$ was avoided in the initial configuration. To speed up the calculations, a cellindexing method was used \cite{allen1999computer}. Here, the number of cells was adjusted to the size of the simulation area and the particle densities. 

To describe the onset of collective motion and the degree of ordering of the particle velocity vectors, we define the usual {\it global} order parameters. On the one hand, polar order is characterized by a vector $\mathbf{P}_j$ with components
\begin{eqnarray}
P_{j,x} &=& \frac{1}{N_j}\sum_{k_j}\cos\theta_{k_j}, \\
P_{j,y} &=& \frac{1}{N_j}\sum_{k_j}\sin\theta_{k_j},
\end{eqnarray}
where the sum is over all particles $k_j$ of species $j$ ($j=1,2$) and $N_2=N-N_1$ equals the number of particles of species $2$. The degree of global polar order is then obtained as the magnitude of this vector, 
\begin{equation}\label{defP}
P_j=\sqrt{P_{j,x}^2+P_{j,y}^2}, \qquad j=1,2. 
\end{equation}
On the other hand, nematic order is described by a traceless symmetric tensor $\mathbf{Q}_j$ with components
\begin{eqnarray}
Q_{j,xx} &=& \frac{1}{N_j}\sum_{k_j}\left[\cos^2\!\theta_{k_j}-\frac{1}{2}\right], \\
Q_{j,xy}=Q_{j,yx} &=& \frac{1}{N_j}\sum_{k_j}\cos\theta_{k_j}\sin\theta_{k_j}, \\
Q_{j,yy} &=& \frac{1}{N_j}\sum_{k_j}\left[\sin^2\!\theta_{k_j}-\frac{1}{2}\right], 
\end{eqnarray}
which implies $Q_{j,xx}=-Q_{j,yy}$ ($j=1,2$). We obtain the degree of global nematic order as
\begin{equation}\label{defS}
S_j = 2\sqrt{Q_{j,xx}^2+Q_{j,xy}^2} = 2\sqrt{Q_{j,xy}^2+Q_{j,yy}^2}.
\end{equation}

In general, spatial heterogeneities emerge close to the onset of collective motion as noted in the Introduction. To first focus on the orientational order itself and reduce the effect of possible spatial heterogeneities, we studied relatively small systems. These were of size $L_x=L_y=0.83$ with $N_1=N_2=117$ particles. We varied the coupling parameter $g$ and determined the degrees of orientational order. These were averaged over $100$ independent runs that started from different initial conditions. 

To study the occurrence of spatial heterogeneities above onset, we turned to larger system sizes. We report results for quadratic simulation areas of $L_x=L_y=10$ and $N_1=N_2=17000$.

\subsection{Polar and antiparallel alignment}

\begin{figure}
\includegraphics[width=8.5cm]{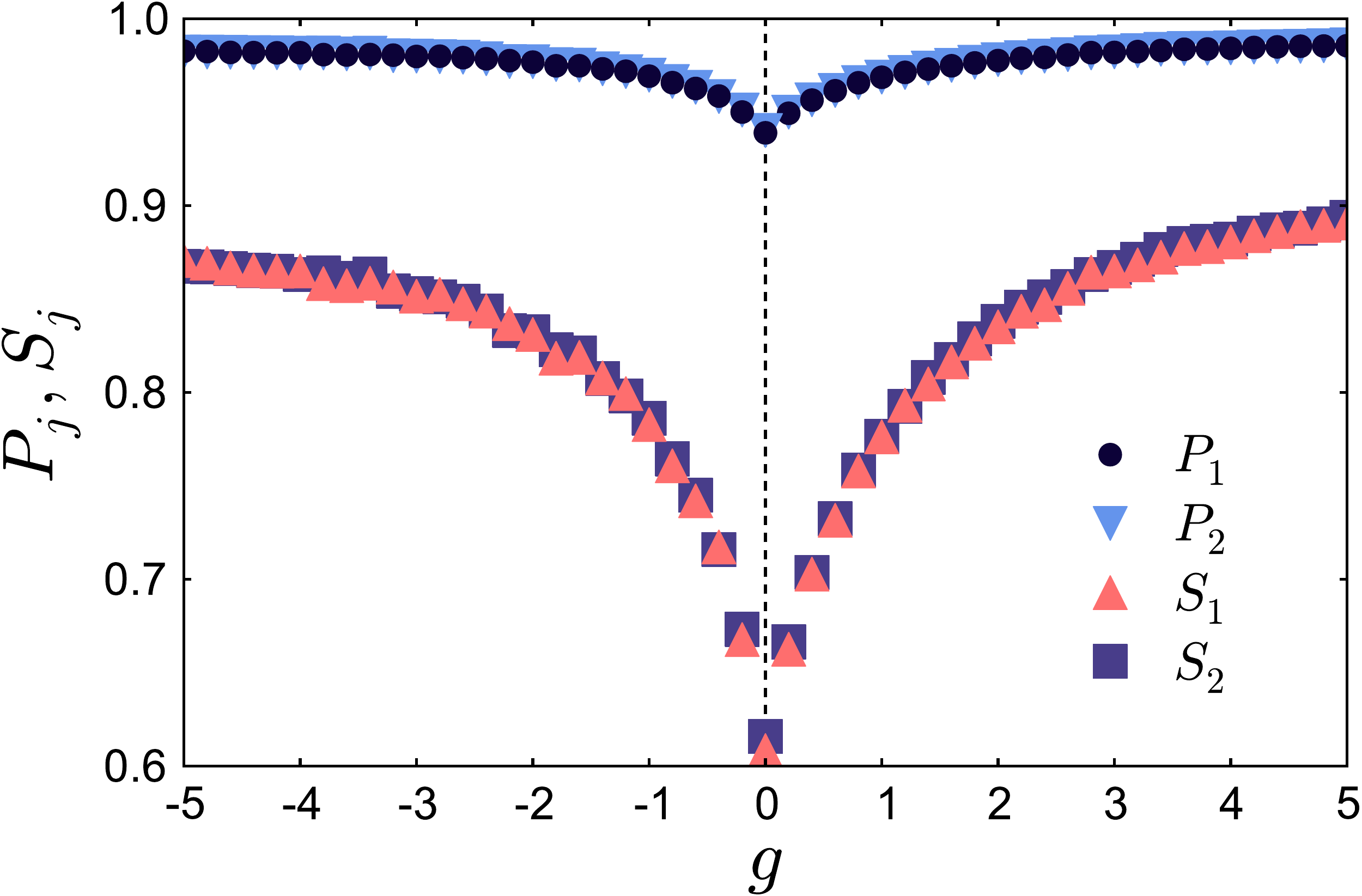}
\caption{(Color online) Polar and nematic degrees of orientational order $P_j$ and $S_j$ as a function of the coupling parameter $g$ for preferred polar ($g>0$) and antiparallel ($g<0$) alignment between the two species $j=1,2$. Approximately, the two cases are symmetric to each other, with slightly smaller values on the antiparallel ($g<0$) side for higher values of $|g|$. The system size was relatively small with $L_x=L_y=0.83$ and $N_1=N_2=117$ to reduce the influence of spatial heterogeneities. Results are averaged over $100$ independent runs. Other parameter values were $u_1=u_2=1$, $g_1=g_2=1$, and $D_1=D_2=1$ during the corresponding particle simulations.} 
\label{gN234par}
\end{figure}

The polar and nematic degrees of orientational order obtained for the small system sizes are depicted in Fig.~\ref{gN234par} for each species. 
Due to the inherent symmetry the two species show approximately the same behavior. For each particle species separately, the density is above the critical threshold density for the onset of collective motion. Therefore, we obtain nonzero order parameters already for vanishing coupling between the two species at $g=0$. 

We find polar orientational order within each species $P_j\neq0$ that also leads to a nonzero value $S_j\neq0$, $j=1,2$. With increasing magnitude of coupling $|g|$ between the two species the orientational order increases. The two species support each other in orientational ordering. Approximately, the curves for $g<0$ and $g>0$ are symmetric to each other with respect to the line at $g=0$. At higher values of $|g|$ the values for $g<0$ are slightly lower than the ones for $g>0$. This seems natural since for preferred polar alignment the interacting pairs of particles of different species move into the same direction. They have a longer time of interaction compared to the antiparallel case, where they only meet at an instant. 

\begin{figure}
\includegraphics[width=8.5cm]{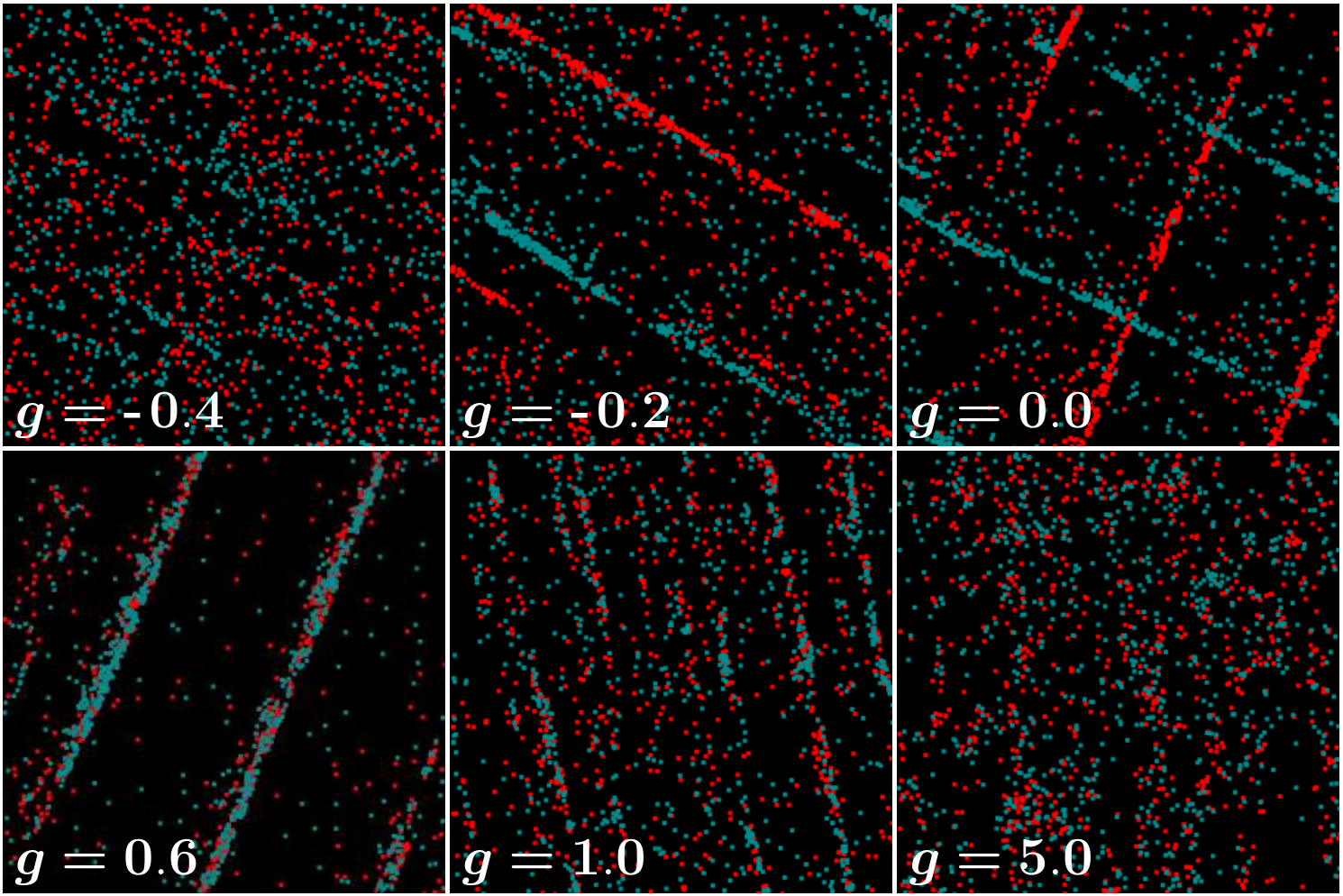}
\caption{(Color online) Snapshots of particle simulations for densities above the onset of collective motion for preferred polar ($g>0$) and antiparallel ($g<0$) alignment between different particle species. The system size was $L_x=L_y=10$ with $N_1=N_2=17000$ particles, the snapshots were taken after $30000$ iterations of time step $dt=0.1$. Other parameter values were $u_1=u_2=1$, $g_1=g_2=1$, and $D_1=D_2=1$. The two species are indicated by red (darker in grayscale) and turquoise (brighter in grayscale), and only every $20th$ particle position is marked. Clearly, an increasing magnitude of coupling $|g|$ between the two species decreases the spatial heterogeneity in the system.} 
\label{particle_stripes_par}
\end{figure}

For larger system sizes, spatial heterogeneities develop above the onset of collective motion. Example snapshots are shown in Fig.~\ref{particle_stripes_par}. 

We start our discussion with the snapshot for $g=0.0$. In this case the two particle species are decoupled and form independent subsystems. Both particle densities themselves are above (but not too far above) the critical single-particle system density. Therefore, we observe the emergence of the traveling stripes or fronts as it was found for the single-particle case \cite{chate2008collective,chate2008modeling,peruani2011polar,mishra2010fluctuations}. The stripes move perpendicularly to the direction of their elongation with nearly the single-particle speed. It is by accident that the stripes for different particles are oriented almost perpendicularly to each other in this picture. 

When we decrease the coupling parameter $g$ to negative values, antiparallel alignment between the two particle species is preferred. In the snapshot for $g=-0.2$ we can see that the stripes are now oriented parallel to each other. The two stripes for different species feature opposite directions of collective motion. Therefore, from time to time they penetrate through each other satisfying the antiparallel alignment rule. 

On further decreasing $g$, the spatially heterogeneous stripe order is destroyed as can be seen from the picture for $g=-0.4$, and the system becomes more spatially homogeneous. This process takes place in the regime where the degrees of orientational order still strongly increase with $|g|$. 

When we increase $g$ to positive values, we induce polar alignment between the two particle species. Now the stripes for the two particle species tend to move into the same direction. As a result, particles of different species mix and form one large stripe, as shown by the snapshot for $g=0.6$. Again, this compound stripe moves perpendicularly to the direction of its elongation. 

Further increasing $g$ destroys the spatial heterogeneity also in the case of polar alignment. In the picture of $g=1.0$, stripe-like residues are still visible, but they do not form a compound object as for the case of $g=0.6$. For $g=5.0$ the thin stripes are not visible any more. 

In summary, we found that an increasing magnitude of coupling interaction $|g|$ reduces the spatial heterogeneity. We can understand this by analogy to the single-particle case. For the latter, it was reported that stripes only persist above but close to the onset of collective motion \cite{chate2008collective,chate2008modeling,peruani2011polar}. Far above onset, the single-particle systems were found to become spatially homogeneous again. 

Effectively, we observe the same phenomenon in our systems when we increase the coupling between the two species $|g|$. It is easiest to see this in the polar case. For the value $g=1.0$, we effectively obtain a single-species system of twice the density as for $g=0.0$. Therefore, we have effectively increased the particle density to far above its threshold value for the onset of collective motion. This significantly reduces the spatially heterogeneous ordering into stripes. The latter becomes obvious when we compare the snapshot for $g=1.0$ to the cases of $g=0.0$ and $g=0.6$. 

Apparently, the effect is much stronger in the case of preferred antiparallel alignment between different particle species when compared to polar alignment. This is expected for the following reason. In the polar case, where the two species form one compound stripe, the interaction between particles can be effectively reduced. The latter is possible by an increased elongation of the stripe or by splitting into several stripes. Since all of these objects move into the same direction, they do not meet each other and consequently the frequency of interaction is lower. In the antiparallel case, the stripes for different species move into antiparallel directions and from time to time must penetrate through each other. When these events occur, the density of particles interacting is doubled within the stripes and therefore remains high above the threshold density for the onset of collective motion.

\subsection{Perpendicular alignment}

\begin{figure}
\includegraphics[width=8.5cm]{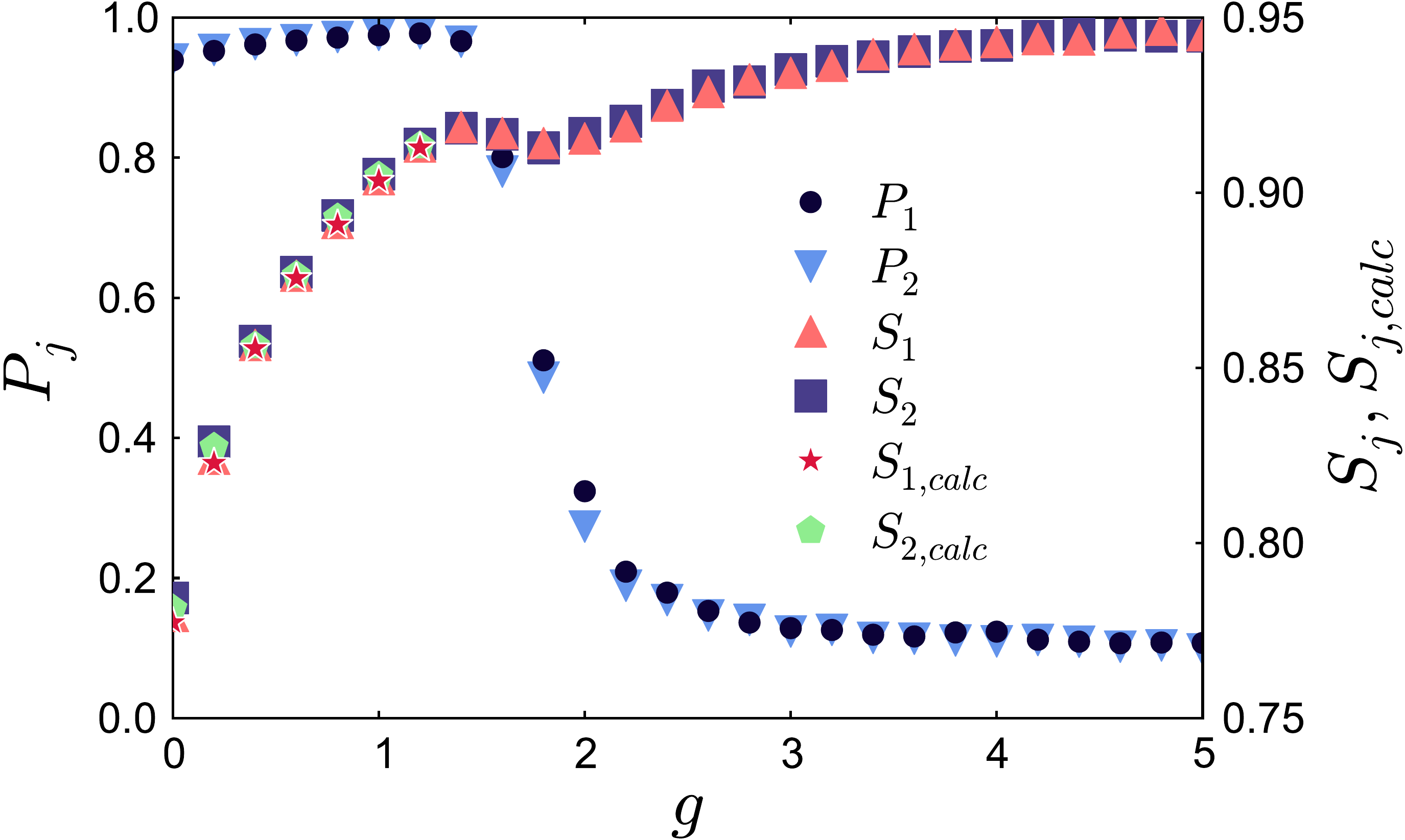}
\caption{(Color online) Polar and nematic degrees of orientational order $P_j$ and $S_j$ as a function of the coupling parameter $g$ for preferred perpendicular alignment between the two species $j=1,2$. The values $S_{j,calc}$ were calculated solely from the magnitude of the corresponding $P_j$ values. A transition from predominantly polar to non-polar nematic order is obvious around $g=2$. Technical details as given by the caption of Fig.~\ref{gN234par}.} 
\label{gN234}
\end{figure}

Again, we first focus on the small systems, where the role of spatial heterogeneities is reduced. As expected from section \ref{linstabanal}, this case of perpendicular alignment is qualitatively different and richer in phenomena than the corresponding parallel case. The results for the degrees of polar and nematic order are depicted in Fig.~\ref{gN234}. 
We can see that the two particle species show quantitatively the same behavior, as expected for reasons of symmetry. 

For small values of $g$ polar order dominates within each species. Here, the magnitudes of the degrees of polar orientational order $P_j$ ($j=1,2$) are close to $1$. As we can see from Fig.~\ref{gN234}, the values of $P_j$ slightly increase with increasing $g$. Apparently the scattering between particles of different species enhances the polar ordering within each species for small values of $g$. In addition, we could show that the nonzero degree of nematic order in this regime is only due to polar orientational order. For this purpose, we assumed a Gaussian distribution of the orientational angles $\theta_{k_j}$ around their mean value for each species. This assumption is corroborated by the observations from the simulations. For each value of $g$, the Gaussian distribution is completely determined via the corresponding value of $P_j$. We then use this Gaussian angular distribution to calculate the degree of nematic orientational order, which we call $S_{j,calc}$. As depicted in Fig.~\ref{gN234} the values thus obtained are identical with those of $S_j$ that were extracted directly from the simulations. 

Around the value $g=2$ polar orientational order breaks down. Eq.~(\ref{gthresh}) predicts a threshold value of $g$ at which the mode $n=2$ can become unstable first. Indeed, we observe that the degrees of nematic orientational order $S_j$ further increase with increasing values of $g\geq 2$. The small dip in $S_j$ reflects the drop in polar orientational order, which is then compensated by truly nematic (mode $n=2$) orientational order. 

\begin{figure}
\includegraphics[width=8.5cm]{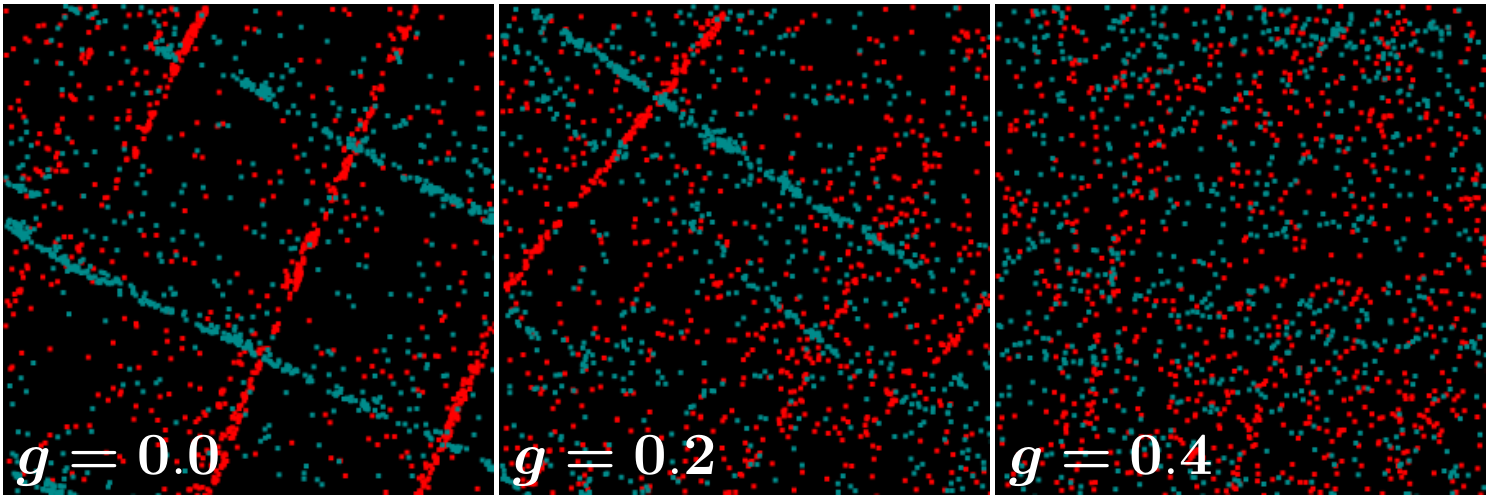}
\caption{(Color online) Snapshots of particle simulations for densities above the onset of collective motion for preferred perpendicular alignment between different particle species. Technical details are the same as given by the caption of Fig.~\ref{particle_stripes_par}. Again, an increasing magnitude of coupling $g$ between the two species decreases spatial heterogeneity in the system.} 
\label{particle_stripes_perp}
\end{figure}

For the larger system sizes that feature spatial heterogeneities in the form of stripe textures we depict example snapshots in Fig.~\ref{particle_stripes_perp}. 
The case is similar to the one of preferred antiparallel alignment that was shown in Fig.~\ref{particle_stripes_par} for $g>0$. 

The state of decoupled species at $g=0.0$ is identical to the one in Fig.~\ref{particle_stripes_par}. Already for $g=0.2$ the spatial ordering into stripes is noticeably reduced. The stripe objects move approximately perpendicularly to each other. For $g=0.4$ the stripes have dissolved. 

As in the antiparallel case, the three snapshots fall into the regime where the overall degrees of orientational order still strongly increase with the magnitude of $g$. In contrast to the antiparallel case, the stripes never completely overlap along their elongation. However, they are always in contact at one crossing intersection. There, the density of interacting particles is again doubled and therefore far above the onset value for collective motion.

\section{Numerical solutions of the Fokker-Planck equations}\label{numFP}

The results of the particle simulations compare well to the numerical results obtained from the continuum equations. To solve Eqs.~(\ref{eq1}) and (\ref{eq2}) numerically, we used a finite differencing scheme. In analogy to the particle simulations, the spatial calculation grid was quadratic of size $N_x\times N_y$ with periodic boundary conditions. $N_{\vartheta}$ is the number of possible angular orientations considered. It was chosen such that the convolution integrals in the angular distributions could be efficiently evaluated via fast Fourier transforms \cite{press1992numerical}. We found that a first order upwind scheme for the first order spatial derivatives reproduces well the physical properties of the system. The time step must be small enough to conserve the overall particle densities. 

As for the particle simulations, we focus on two cases. First, we investigate spatially homogeneous solutions by setting $N_x=N_y=1$ ($N_{\vartheta}=128$). After that, the influence of spatial degrees of freedom is taken into account, where we mostly used $N_x=N_y=32$ ($N_{\vartheta}=32$). The parameters were set as for the particle simulations, except for the velocities ($u_1=u_2=0.1$, $g_1=g_2=1$, $D_1=D_2=1$), and we varied the interaction parameter $g$. For our choice of parameter values, we obtain from Eq.~(\ref{critrhosep}) a critical single-species system density of $\rho_{j0}^{*,1}=1$ ($j=1,2$). We report results for three characteristic scenarios: (a) both values of the mean densities $\rho_{j0}$ ($j=1,2$) are above the critical single-species system densities ($\rho_{10}=\rho_{20}=1.5$); (b) one of the mean densities is above the critical single-species system density and one is below (we consider $\rho_{10}=1.5$, $\rho_{20}=0.5$ for spatially homogeneous solutions and $\rho_{10}=1.1$, $\rho_{20}=0.5$ for spatially inhomogeneous ones); and (c) both mean densities are below the single-species system density ($\rho_{10}=\rho_{20}=0.5$). For each value of $g$, we initialized the densities on the $N_x\times N_y\times N_{\vartheta}$ sized calculation grid by the mean densities plus a random number of Gaussian distribution and small amplitude. 

In the continuum picture, we obtain the {\it local} orientational order parameters by taking the moments of the densities $\rho_1(\mathbf{r},\theta,t)$ and $\rho_2(\mathbf{r},\theta,t)$ with respect to the angular distributions, 
\begin{eqnarray}
c_j(\mathbf{r},t) &=& \int_{0}^{2\pi}\rho_j(\mathbf{r},\theta,t)d\theta, \label{eqc} \\[.1cm]
c_j(\mathbf{r},t)\mathbf{P}_j(\mathbf{r},t) &=& \int_{0}^{2\pi}
  \left(\begin{array}{c} \cos\theta \\ \sin\theta \end{array}\right)
  \rho_j(\mathbf{r},\theta,t)d\theta, \label{eqcP} \\[.1cm]
c_j(\mathbf{r},t)\mathbf{Q}_j(\mathbf{r},t) &=& \int_{0}^{2\pi}
  \left(\begin{array}{cc} \cos^2\!\theta-\frac{1}{2} & \cos\theta\sin\theta \\ 
  \cos\theta\sin\theta & \sin^2\!\theta-\frac{1}{2} \end{array}\right) \nonumber\\[.1cm]
&& {}\qquad\qquad\qquad\times \rho_j(\mathbf{r},\theta,t)d\theta, \label{eqcQ}
\end{eqnarray}
$j=1,2$. Here, $c_j(\mathbf{r},t)$ gives the {\it local} particle number density. $\mathbf{P}_j(\mathbf{r},t)$ corresponds to the {\it local} polar alignment vector, whereas $\mathbf{Q}_j(\mathbf{r},t)$ is the {\it local} nematic order parameter tensor for each species $j=1,2$. At each position $\mathbf{r}$, the {\it local} degrees of orientational order follow in analogy to Eqs.~(\ref{defP}) and (\ref{defS}). To obtain {\it global} degrees of orientational order, we took the spatial averages over the system size. Discretized versions of these definitions were used for the numerical implementation.

\subsection{Polar and antiparallel alignment}

\begin{figure}
\includegraphics[width=8.5cm]{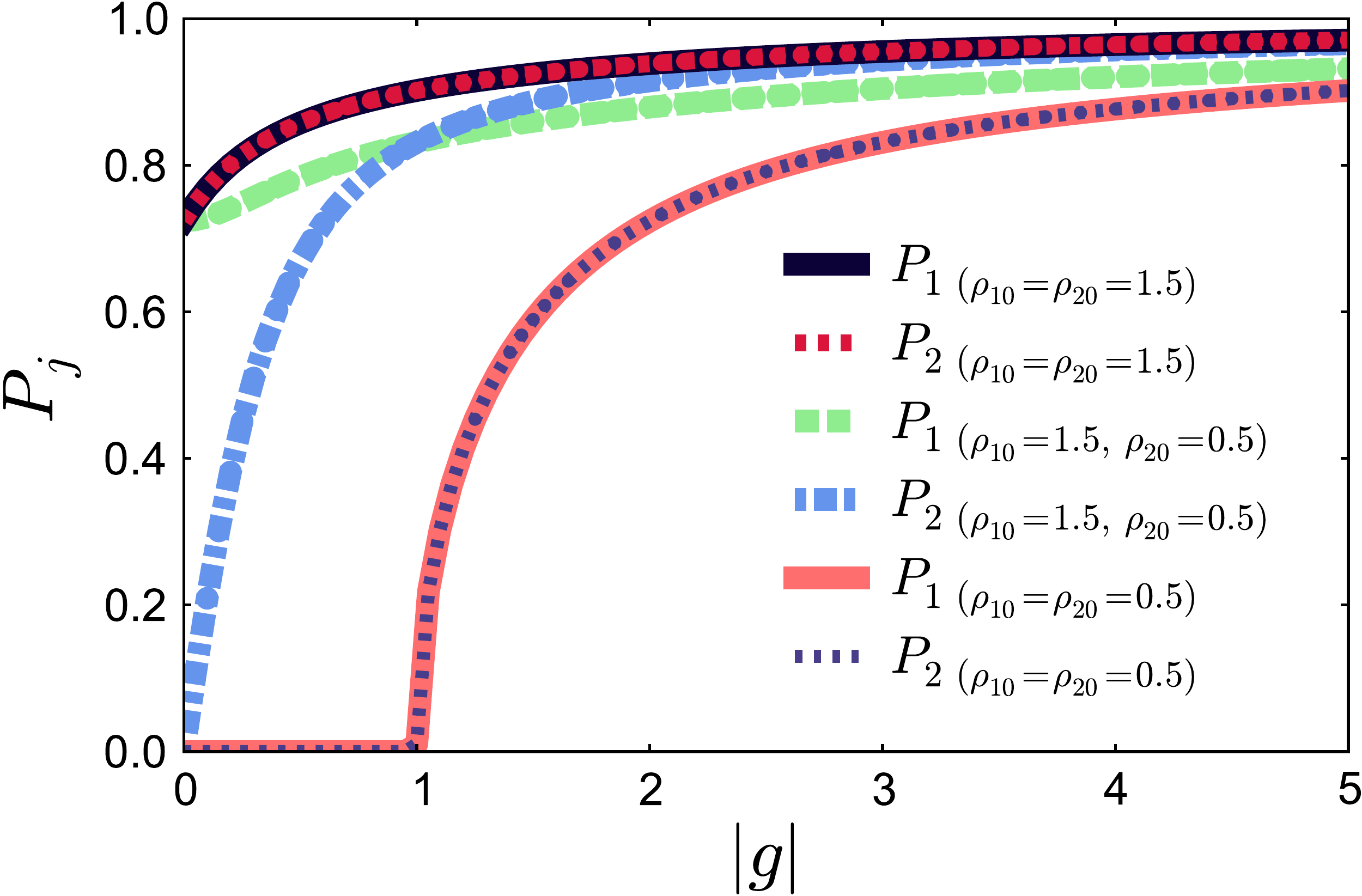}
\caption{(Color online) Polar degrees of orientational order $P_j$ for the two species $j=1,2$ as a function of the coupling parameter $g$. The results were obtained from spatially homogeneous numerical solutions of the Fokker-Planck equations for preferred antiparallel orientational order between the two species. Different mean particle densities were considered. Other parameter values were $u_1=u_2=0.1$, $g_1=g_2=1$, and $D_1=D_2=1$.} 
\label{FPantiparhom}
\end{figure}

First, we confine ourselves to spatially homogeneous solutions. We show characteristic results for preferred antiparallel alignment between particles of different species in Fig.~\ref{FPantiparhom} as a function of $|g|$ (results for polar alignment follow approximately analogously for $g$-values of opposite sign, compare also Fig.~\ref{gN234par}). 
The degree of nematic order is nonzero due to polar orientational order and not depicted in the figure. 

For $\rho_{10}=\rho_{20}=1.5$ both species densities are above the critical one-component density. There is collective motion already without coupling at $g=0$. The relative angular orientation between the two species is not yet fixed, however. For nonzero values of $|g|$, the polar or antiparallel orientation of the two species velocities sets in. The two species support each other in orientational ordering with increasing values of $g$. 

The asymmetric case, where $\rho_{10}=1.5$ is above and $\rho_{20}=0.5$ is below the threshold density, is interesting for the following reason. For both species, orientational order increases due to the coupling to the other species with increasing $|g|$-values. For the given density values, $P_2$ starts to grow already at $g=0$. Remarkably, the degree of orientational order of the more diluted species $P_2$ even exceeds the one of the denser species $P_1$ for strong coupling between the two species. Surprisingly, for high values of $|g|$, it is $P_2$ that asymptotically approaches the degree of order that was reached for $\rho_{10}=\rho_{20}=1.5$. In contrary, $P_1$ asymptotically approaches the lower degree of order reached for $\rho_{10}=\rho_{20}=0.5$ (see below), despite the higher mean density $\rho_{10}=3\rho_{20}$. So $P_1$ and $P_2$ behave oppositely to what would be expected from the corresponding mean densities. However, we note that, at high values of $|g|$, orientational order of one species is predominantly achieved by interactions with the other species. In this way, it is the density of the other species that determines the asymptotic degree of ordering. 

In the case of $\rho_{10}=\rho_{20}=0.5$ there is no collective motion for $g=0$. Only at $|g|=1$ collective motion sets in for both species simultaneously. At this value, the two species form an effectively single-component system, so that the densities add up to an effective density $\rho_{10}+\rho_{20}=1$. This value is the threshold density for the onset of collective motion in a corresponding single-component system [compare, e.g., Eq.~(\ref{critrhosep})]. 

\begin{figure}
\includegraphics[width=5.67cm]{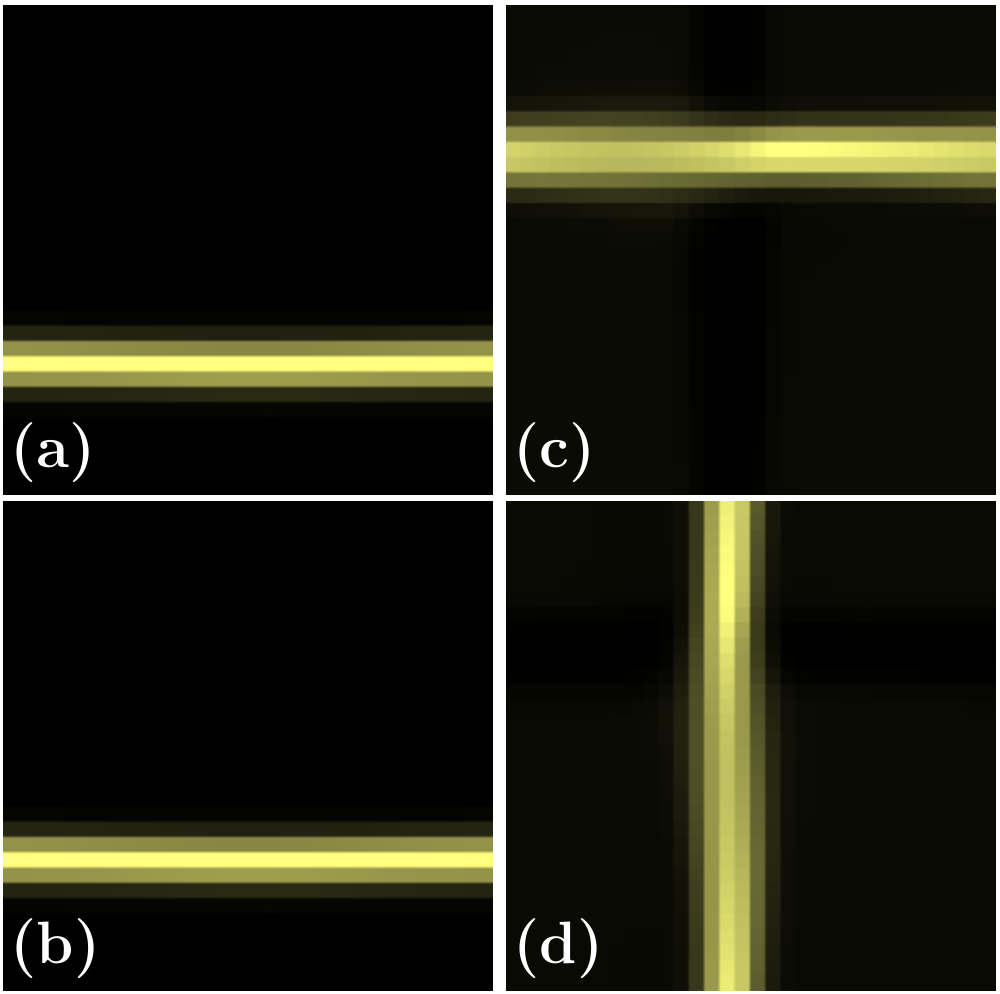}
\caption{(Color online) Spatial density distributions obtained from numerical solutions of the Fokker-Planck equations for mean densities $\rho_{10}=\rho_{20}=1.5$. Upper panels [(a) and (c)] correspond to species $1$, lower ones [(b) and (d)] to species $2$. Left panels [(a) and (b)] follow from preferred polar alignment between the different species with $g=0.4$ and show parallel orientation of the resulting stripes. Right panels [(c) and (d)] follow from preferred perpendicular alignment with $g=0.6$ and feature perpendicular orientation. Other parameter values were $u_1=u_2=0.1$, $g_1=g_2=1$, and $D_1=D_2=1$. The numerical grid size was $N_x\times N_y=32^2$ lattice points of distance $dx=0.5$ with $N_{\vartheta}=32$ angular orientations each, and the equations were iterated $3\times10^6$ times with step size $dt=0.001$. Brightness increases with density and has been rescaled to maximize the spatial density contrast.} 
\label{stripes1515}
\end{figure}

Second, we investigated the spatial heterogeneities that appear for the larger system sizes. Qualitatively, our results obtained from the particle simulations in the previous section are confirmed. They correspond to the case $\rho_{10}=\rho_{20}=1.5$, in which both mean densities are above the single-species threshold density. Spatial heterogeneities in the form of stripes appear in the density profiles. An example is shown in Fig.~\ref{stripes1515}~(a) and (b). 

Again, we find that an increasing magnitude of the coupling parameter $|g|$ dissolves the stripes. As for the particle simulations, the necessary values for $|g|$ are much smaller for antiparallel than for polar alignment between different species. The numerical solutions of the Fokker-Planck equations offer a simple method to determine this value of $|g|$: the difference between the largest and the smallest density value decays to zero when $|g|$ destroys the spatial heterogeneity. We find a value of $|g|\approx 4.8$ in the polar and $|g|<0.2$ in the antiparallel case. In contrast to the particle simulations, we often observe the velocity vectors to align along the stripe direction for antiparallel interactions. The density profiles of the two stripes of different species then are stationary and overlap. This increases the interaction time between the two species densities. 

The asymmetric case of $\rho_{10}=1.1$ and $\rho_{20}=0.5$ shows interesting phenomena. For the majority species $1$ the mean density is above onset. Consequently collective motion sets in and spatial heterogeneities occur. Through the interaction of strength $g$, collective motion can also be induced in the minority species $2$. However, this happens only at positions where the density $c_1(\mathbf{r},t)$ is high enough. 

\begin{figure}
\includegraphics[width=8.5cm]{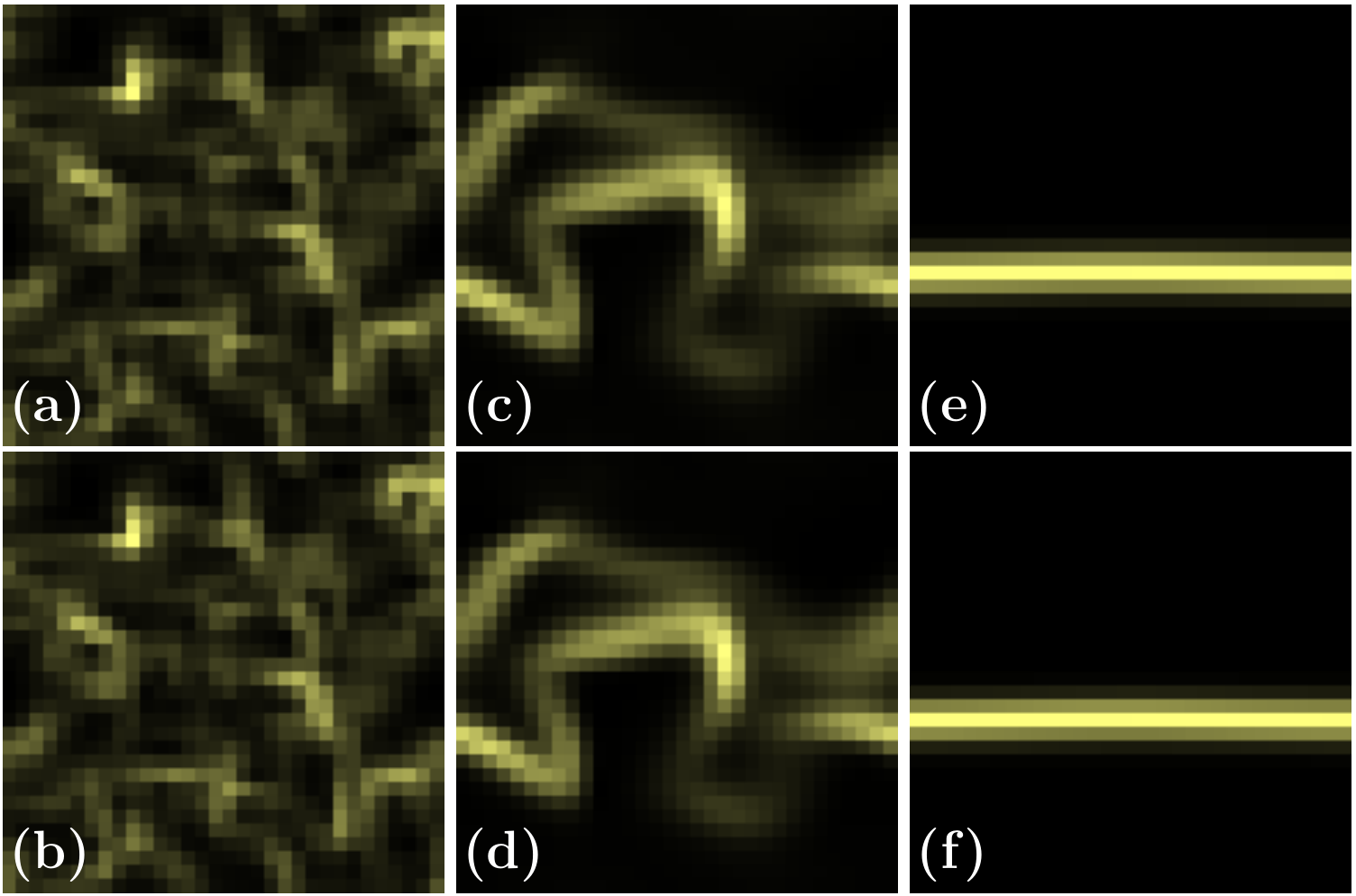}
\caption{(Color online) Evolution of the spatial density distribution obtained from numerical solutions of the Fokker-Planck equations for mean densities $\rho_{10}=1.1$ and $\rho_{20}=0.5$ in the case of polar alignment. The coupling strength between the two species was $g=2.2$. Upper panels [(a), (c), and (e)] correspond to species $1$, lower ones [(b), (d), and (f)] to species $2$. Shown snapshots were taken at the following numerical times: (a) and (b) $10$; (c) and (d) $50$; (e) and (f) $3000$. At each depicted instant the density map for species $2$ is a copy of the map for species $1$. Further technical details as given by the caption of Fig.~\ref{stripes1515}.} 
\label{copydensitymap}
\end{figure}

For polar alignment interactions, Fig.~\ref{copydensitymap}~(e) shows the final density profile of a stripe of species $1$. 
Within the stripe region, $c_1(\mathbf{r},t)$ is high and induces collective motion in species $2$. At these locations, the material of both species propels into the same direction. As a result, spots of high density $c_2(\mathbf{r},t)$ follow the ones of high density $c_1(\mathbf{r},t)$. In this way, the density map of $c_2(\mathbf{r},t)$ depicted in Fig.~\ref{copydensitymap}~(f) becomes a copy of the one of $c_1(\mathbf{r},t)$. Consequently, an induced overlapping stripe of species $2$ is generated with polar alignment of the collective velocity vector. These statements even hold at early times of the ordering process when the stripe textures have not yet developed. An example is given by the time series in Fig.~\ref{copydensitymap}~(a)--(f).

\begin{figure}
\includegraphics[width=5.67cm]{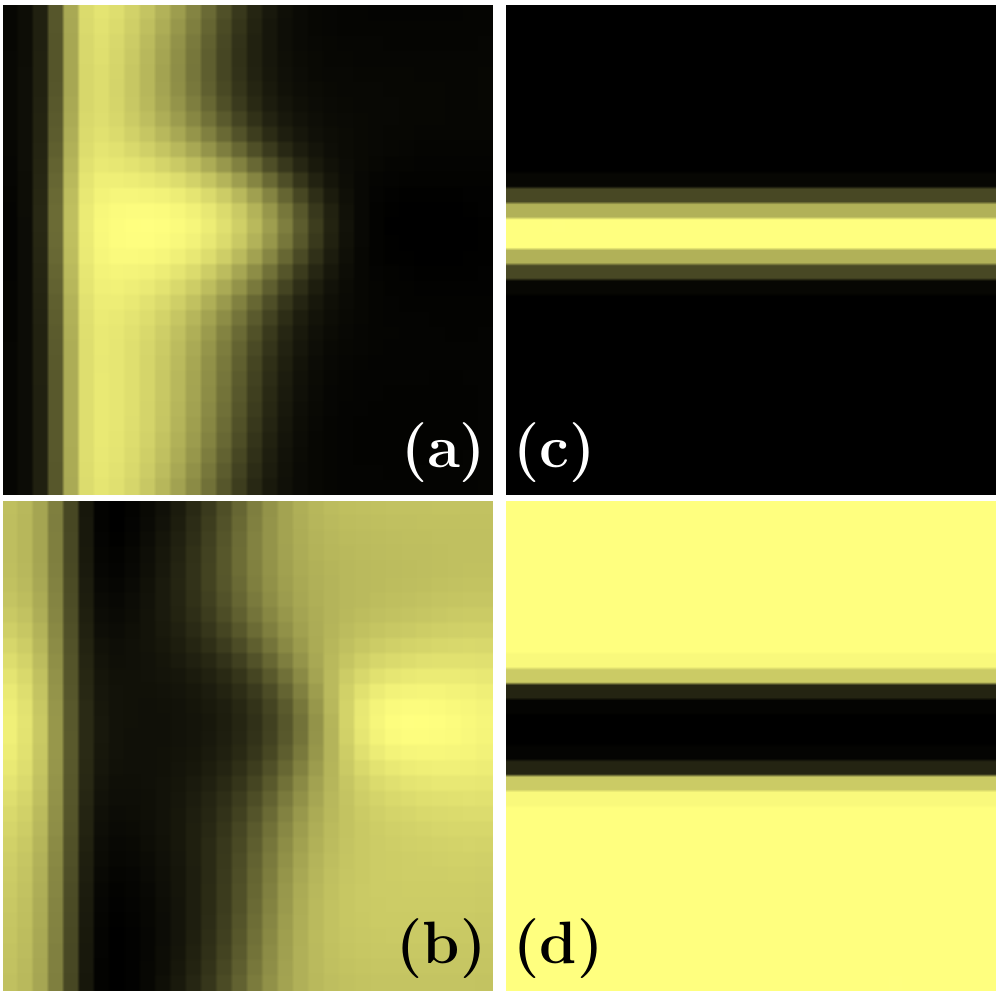}
\caption{(Color online) Spatial density distributions obtained from numerical solutions of the Fokker-Planck equations for mean densities $\rho_{10}=1.1$ and $\rho_{20}=0.5$. Upper panels [(a) and (c)] correspond to species $1$, lower ones [(b) and (d)] to species $2$. Left panels [(a) and (b)] follow from preferred antiparallel alignment between the different species with $g=-0.2$. The stripe in panel (a) travels to the left and features the typical sharp front and fuzzy tail. Material of species $2$ is advected through the stripe in opposite direction. Right panels [(c) and (d)] follow from preferred perpendicular alignment with $g=2.2$. Again the flow field in panel (c) is oriented to the left. Material of species $2$ is expelled from the stripe region to the top and bottom. Both cases lead to inverted spatial density profiles for species $2$. Further technical details as given by the caption of Fig.~\ref{stripes1515}.} 
\label{stripesantiparinverted}
\end{figure}

It is interesting to note, however, that for the antiparallel case the density profiles are inverted. This is illustrated by Fig.~\ref{stripesantiparinverted}~(a) and (b), where a stripe of species $1$ is moving to the left. The motion to the left leads to the sharp front and the fuzzy tail of the stripe in panel (a). 
In the region of the stripe, the density $c_1(\mathbf{r},t)$ is high so that it can induce collective motion in species $2$. This happens via the antiparallel alignment interaction $g<0$. Consequently, the vector of collective motion of species $2$ points into the opposite direction, i.e.~to the right. In this way, material of species $2$ at the head of the moving stripe is ``pumped'' through the stripe of species $1$. Behind and outside the stripe of species $1$, the density $c_1(\mathbf{r},t)$ is so low that it cannot induce effective collective motion in species $2$ any more. Thus the material of species $2$ is not advected once it has passed the stripe. It gathers behind the stripe area. In effect, this leads to the inverted stripe density profile shown in Fig.~\ref{stripesantiparinverted}~(b). 

As in the spatially homogeneous case, collective motion in the system $\rho_{10}=\rho_{20}=0.5$ sets in at values $|g|\geq 1$. Stripes develop above onset.

\subsection{Perpendicular alignment}

\begin{figure}
\includegraphics[width=8.5cm]{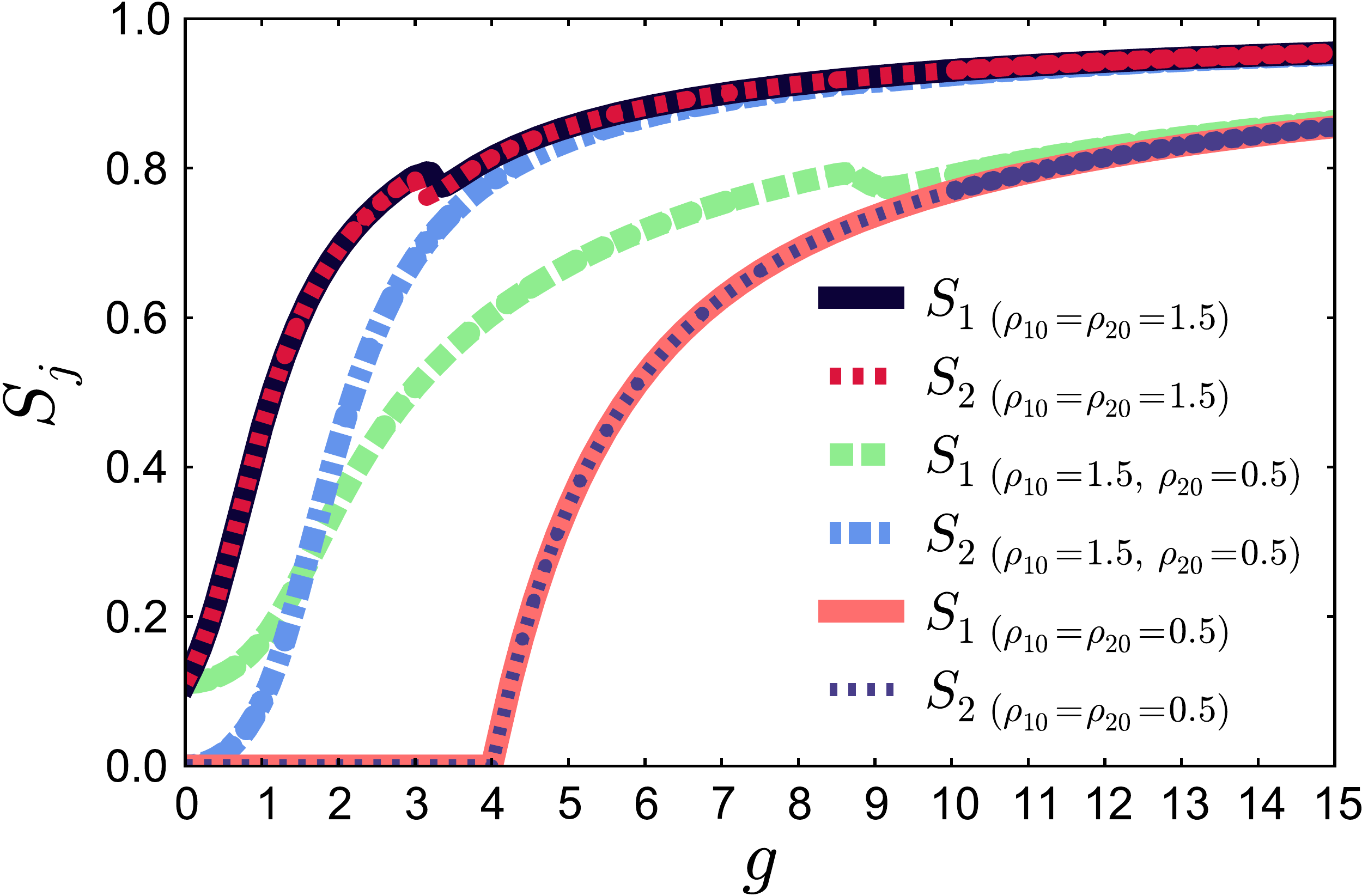}
\caption{(Color online) Nematic degrees of orientational order $S_j$ for the two species $j=1,2$ as a function of the coupling parameter $g$. The results were obtained from spatially homogeneous numerical solutions of the Fokker-Planck equations for preferred perpendicular orientational order between the two species. Different mean particle densities were considered. Other parameter values were $u_1=u_2=0.1$, $g_1=g_2=1$, and $D_1=D_2=1$.} 
\label{FPperphom}
\end{figure}

For preferred perpendicular alignment between particles of different species we again focus on spatially homogeneous solutions first. Typical results are depicted in Fig.~\ref{FPperphom}. 
We only show the degree of nematic order $S_j$ ($j=1,2$). Degrees of polar orientational order remain negligibly small, $P_j\approx0$, for corresponding mean densities $\rho_{j0}=0.5$. If the mean density is $\rho_{0j}=1.5$, the $P_j$ curve has qualitatively the same shape as the ones in Fig.~\ref{gN234}, with the strong descent located at the dip of the respective curve for $S_j$ in Fig.~\ref{FPperphom}. 

At mean density values $\rho_{10}=\rho_{20}=1.5$ collective motion of polar order dominates at low coupling strength $g$. This polar order breaks down where we find the dips in the curves of $S_j$ ($j=1,2$) in Fig.~\ref{FPperphom}. In contrast to the particle picture, where $P_j\approx0.1$ after this transition (see Fig.~\ref{gN234}), we now find values close to zero. At higher values of $g$, truly nematic order dominates. Our linear stability analysis from section \ref{linstabanal} does not provide a quantitative measure for the location of the transition because of the nonzero amplitudes of the polar order parameters below the transition. Differences when compared to the particle simulations illustrate the idealized character of the mean field continuum approach, where finite interaction radii and free paths between particle interactions were not taken into account. 

The asymmetric case of $\rho_{10}=1.5$ and $\rho_{20}=0.5$ features the same interesting effect as in the previous subsection for antiparallel alignment. Species $1$ shows collective motion already at zero coupling $g=0$. Polar orientational order dominates and increases with increasing coupling strength. The polar orientational order breaks down at the location of the dip in the $S_1$ curve in Fig.~\ref{FPperphom}. After that, truly nematic order prevails. $S_1$ then asymptotically approaches the curves of $S_j$ for $\rho_{10}=\rho_{20}=0.5$ (see below) although here $\rho_{10}=1.5$. In contrast, orientational order for species $2$ increases from zero with increasing coupling strength. Polar order never plays a significant role. The value of $S_2$ asymptotically approaches the value of $S_j$ for $\rho_{10}=\rho_{20}=1.5$, despite the fact that $\rho_{20}=0.5$. In analogy to the previous subsection, here $S_1$ and $S_2$ behave oppositely to what would be expected from the corresponding mean densities. Again, it is the density of the other species that determines the asymptotic degree of ordering of one of the two species. This is because at high values of $g$ orientational order of one species is predominantly achieved by interactions with the other species, and not by interactions with particles of the same species. 

In the case of $\rho_{10}=\rho_{20}=0.5$, collective motion sets in for both species simultaneously at $g=4$. This value is predicted by the linear stability analysis via Eq.~(\ref{mod2instab}). Above this threshold, truly nematic orientational order dominates. 

When we turn to the larger system sizes, spatial inhomogeneities can arise. For $\rho_{10}=\rho_{20}=1.5$, both mean species densities are above the critical single-species density. Qualitatively we then find similar behavior as for the particle simulations in Fig.~\ref{particle_stripes_perp}. Stripes develop that are oriented perpendicular to each other as illustrated by the example in Fig.~\ref{stripes1515}~(c) and (d). Similarly to the case of antiparallel alignment, the velocity vectors of collective motion were mainly oriented along the stripe direction. Again, increasing the coupling strength between the two species dissolves the stripes. We found that the system turns spatially homogeneous for $g\geq 1$. 

For $\rho_{10}=1.1$ and $\rho_{20}=0.5$ we observe the same effect as for the antiparallel alignment in Fig.~\ref{stripesantiparinverted}. Since the density for species $1$ is above the critical density, collective motion sets in and spatial heterogeneity in the form of a stripe develops. We find these stripes for $g\leq4.8$. At spots of high density $c_1(\mathbf{r},t)$, i.e.~within the stripes, collective motion is induced in species $2$ due to the interaction of strength $g$. An example result is illustrated in Fig.~\ref{stripesantiparinverted}~(c) and (d). In this situation, the velocity vector of species $1$ is oriented along the stripe. Therefore, material of species $2$ is pumped through and out of the stripe. The consequence is a depletion of material of species $2$ within the stripe region and again an inverted density profile for species $2$, as shown by Fig.~\ref{stripesantiparinverted}~(d). 

Finally, when $\rho_{10}=\rho_{20}=0.5$, we recover the critical value of $g=4$ that was already found for the spatially homogeneous case in Fig.~\ref{FPperphom}. Above this value, $S_1$ and $S_2$ become nonzero and collective motion develops. Interestingly, we found that the systems do not turn spatially heterogeneous directly above onset. At the same time, no polar orientational order was detected. Only for values $g\geq 4.8$ we observed stable spatial heterogeneities. The emergence of these spatial heterogeneities, however, was coupled to the development of polar orientational order within each species. This observation is in contrast to the spatially homogeneous solution, where in this regime we found $P_1\approx 0\approx P_2$. It seems that for the case of perpendicular alignment and $\rho_{10}=\rho_{20}=0.5$ spatial heterogeneities are coupled to the evolution of nonzero values of $P_1$ and $P_2$.

\section{Macroscopic continuum equations}\label{macroscopic}

In this section, we derive macroscopic hydrodynamic-like equations from the Fokker-Planck equations (\ref{eq1}) and (\ref{eq2}) and briefly discuss their regime of validity. We obtain the characteristic macroscopic variables by taking the moments of the densities $\rho_1(\mathbf{r},\theta,t)$ and $\rho_2(\mathbf{r},\theta,t)$ with respect to the angular distributions as given by Eqs.~(\ref{eqc})--(\ref{eqcQ}). Assuming one single mass $m_j$ for particles of each species, $c_j(\mathbf{r},t)$ is proportional to the mass density. $\mathbf{P}_j(\mathbf{r},t)$ gives the {\it local} polar alignment vector, whereas $\mathbf{Q}_j(\mathbf{r},t)$ characterizes the {\it local} nematic order for each species $j=1,2$. Since the magnitude of the velocity is fixed for each species, $c_j(\mathbf{r},t)\mathbf{P}_j(\mathbf{r},t)$ is proportional to the momentum density $m_jc_j(\mathbf{r},t)\mathbf{v}_j(\mathbf{r},t)$. Here, the macroscopic velocity field $\mathbf{v}_j(\mathbf{r},t)$ can be obtained by averaging over all velocity vectors of particles located at time $t$ in a surface element at position $\mathbf{r}$. The corresponding particle velocity vectors are given by Eq.~(\ref{updater}). This leads us to the expressions
\begin{equation}
c_j(\mathbf{r},t)\mathbf{v}_j(\mathbf{r},t) = 2u_j\int_{0}^{2\pi}
  \left(\begin{array}{c} \cos\theta \\ \sin\theta \end{array}\right)
  \rho_j(\mathbf{r},\theta,t)d\theta, 
\end{equation}
where $j=1,2$. 

The dynamic equations are derived by taking the moments of Eqs.~(\ref{eq1}) and (\ref{eq2}) as given by the right-hand sides of Eqs.~(\ref{eqc})--(\ref{eqcQ}). Higher order angular moments are not taken into account. We thus expect quantitative deviations of the results obtained from the hydrodynamic-like equations when compared to direct solutions of the Fokker-Planck equations. This becomes more and more severe when densities are significantly higher than the threshold values. The equations that we list contain terms up to second order in the particle densities $c_1$ and $c_2$. 

In both cases of alignment ($a=1$ and $a=-2$), the zeroth moment of Eqs.~(\ref{eq1}) and (\ref{eq2}) leads to the continuity equation for each species, 
\begin{eqnarray}
\frac{\partial c_j(\mathbf{r},t)}{\partial t} 
& = & {}-\nabla\cdot\left[c_j(\mathbf{r},t)\mathbf{v}_j(\mathbf{r},t)\right] \nonumber\\
& = & {}-2u_j\,\nabla\cdot\left[c_j(\mathbf{r},t)\mathbf{P}_j(\mathbf{r},t)\right],
\end{eqnarray}
$j=1,2$. We find qualitative differences for the higher order angular moment equations for the different cases of alignment.

\subsection{Polar and antiparallel alignment}

In the case of $a=1$, Eq.~(\ref{Ugpar}), we derived the following macroscopic equations. We always refer to {\it local} particle densities as well as {\it local} polar and nematic alignment order parameters. For brevity, however, temporal and spatial coordinates $(\mathbf{r},t)$ are not explicitly noted. 

The temporal evolution of the polar alignment vector $\mathbf{P}_1$ is given by  
\begin{eqnarray} \label{c1P1par}
\partial_t (c_1\mathbf{P}_1) & \approx & 
{}-2u_1\left[\nabla\cdot(c_1\mathbf{Q}_1)+\frac{1}{2}\nabla c_1\right] -D_1c_1\mathbf{P}_1 \nonumber\\
&& {}+\frac{g_1}{\pi}c_1^2\left[\frac{1}{2}\mathbf{P}_1-\mathbf{P}_1\cdot\mathbf{Q}_1\right] \nonumber\\
&& {}+\frac{g}{\pi}c_1c_2\left[\frac{1}{2}\mathbf{P}_2-\mathbf{P}_2\cdot\mathbf{Q}_1\right],
\end{eqnarray}
from which the equation for $\mathbf{P}_2$ follows from switching all subscripts 
\begin{eqnarray} \label{c2P2par}
\partial_t (c_2\mathbf{P}_2) & \approx & 
_1\leftrightarrow\hspace{.4pt}_2
\end{eqnarray}
in Eq.~(\ref{c1P1par}).

Similarly, the temporal evolution of the nematic order parameter $\mathbf{Q}_1$ is obtained as 
\begin{eqnarray} \label{c1Q1par}
\partial_t(c_1\mathbf{Q}_1) & \approx & 
\frac{u_1}{2}\left[\nabla\cdot(c_1\mathbf{P}_1)\right]\mathbf{I} \nonumber\\
&& {}-\frac{u_1}{2}\left[\nabla(c_1\mathbf{P}_1)+\left\{\nabla(c_1\mathbf{P}_1)\right\}^T\right] \nonumber\\[.1cm]
&& {}-4D_1c_1\mathbf{Q}_1 \nonumber\\[.1cm]
&& {}+\frac{g_1}{2\pi}c_1^2\left[2\mathbf{P}_1\mathbf{P}_1-\mathbf{P}_1\!\!^2\,\mathbf{I}\right] \nonumber\\
&& {}+\frac{g}{2\pi}c_1c_2\left[\mathbf{P}_1\mathbf{P}_2+\mathbf{P}_2\mathbf{P}_1-(\mathbf{P}_1\cdot\mathbf{P}_2)\,\mathbf{I}\right]. \nonumber\\
&&
\end{eqnarray}
Here, $^T$ denotes the transpose of the superscripted gradient matrix, and $\mathbf{I}$ corresponds to the unity matrix. The equation for $\mathbf{Q}_2$ follows again from switching all subscripts 
\begin{eqnarray} \label{c2Q2par}
\partial_t(c_2\mathbf{Q}_2) & \approx & 
_1\leftrightarrow\hspace{.4pt}_2
\end{eqnarray}
in Eq.~(\ref{c1Q1par}). 

We can see that diffusion tends to reduce both polar and nematic order through the terms proportional to $D_j>0$, $j=1,2$. This term is obtained from the second partial derivative $\partial_{\theta}^2$ in Eqs.~(\ref{eq1}) and (\ref{eq2}). Consequently, it grows quadratically in the angular order of the considered mode. This allows to approximately neglect higher modes of orientational order as mentioned above. 

These equations are now analyzed with respect to the threshold for the onset of collective motion. For the reasons noted in section \ref{linstabanal}, we restrict ourselves to the spatially homogeneous case. I.e.~we neglect the gradient terms in Eqs.~(\ref{c1P1par})-(\ref{c2Q2par}). Close to threshold, the second angular momenta, represented by $\mathbf{Q}_1$ and $\mathbf{Q}_2$, relax faster than the first angular momenta $\mathbf{P}_1$ and $\mathbf{P}_2$. Therefore, we set the partial time derivatives on the left-hand sides of Eqs.~(\ref{c1Q1par}) and (\ref{c2Q2par}) to zero and solve for the stationary spatially homogeneous values of $\mathbf{Q}_1$ and $\mathbf{Q}_2$, 
\begin{eqnarray}
\mathbf{Q}_{1,st} & \approx & \frac{1}{4D_1}\bigg\{\frac{g_1}{2\pi}c_1\left[2\mathbf{P}_1\mathbf{P}_1-\mathbf{P}_1\!\!^2\,\mathbf{I}\right] \nonumber\\
&& {}+\frac{g}{2\pi}c_2\left[\mathbf{P}_1\mathbf{P}_2+\mathbf{P}_2\mathbf{P}_1-(\mathbf{P}_1\cdot\mathbf{P}_2)\,\mathbf{I}\right]\bigg\}, \nonumber\\
&&
\end{eqnarray}
and
\begin{eqnarray}
\mathbf{Q}_{2,st} & \approx & _1\leftrightarrow\hspace{.4pt}_2.
\end{eqnarray}
Inserting into Eqs.~(\ref{c1P1par}) and (\ref{c2P2par}) leads to
\begin{eqnarray}\label{c1P1paradiabatic}
\partial_t(c_1\mathbf{P}_1) & \approx & 
c_1\left(\frac{g_1}{2\pi}c_1-D_1\right)\mathbf{P}_1
+\frac{g}{2\pi}c_1c_2\mathbf{P}_2 \nonumber\\
&& {}-\frac{1}{4D_1}\frac{1}{2\pi^2}c_1\left(g_1c_1\mathbf{P}_1+gc_2\mathbf{P}_2\right)^2\mathbf{P}_1 \nonumber\\
&&
\end{eqnarray}
and
\begin{eqnarray}\label{c2P2paradiabatic}
\partial_t(c_2\mathbf{P}_2) & \approx & _1\leftrightarrow\hspace{.4pt}_2.
\end{eqnarray}
From a linear stability analysis of these equations around the non-ordered state $\mathbf{P}_1=\mathbf{P}_2=\mathbf{0}$, we obtain the same eigenvalues as those in Eq.~(\ref{eigenvalpar}). This leads to the same threshold values for the onset of collective motion as derived in section \ref{linstabanal}. 

Crossing the threshold for the onset of collective motion, it is important to note that the system of Eqs.~(\ref{c1P1paradiabatic}) and (\ref{c2P2paradiabatic}) is not stable for large values of the average densities. This is already the case for the spatially homogeneous single-species scenario that we obtain from Eqs.~(\ref{c1P1paradiabatic}) and (\ref{c2P2paradiabatic}) by setting, for example, $c_2=0$. For a single-species system, the analog to Eq.~(\ref{c1P1paradiabatic}) reads
\begin{eqnarray}
\partial_t(c_1\mathbf{P}_1) & \approx & 
c_1\left(\frac{g_1}{2\pi}c_1-D_1\right)\mathbf{P}_1 \nonumber\\
&& {}-\frac{1}{4D_1}\frac{1}{2\pi^2}g_1^2c_1^3\mathbf{P}_1\!\!^2\mathbf{P}_1. 
\end{eqnarray}
Despite the stabilizing cubic term in $\mathbf{P}_1$, a systematic linear stability analysis shows that the static solution becomes linearly unstable at density values
\begin{equation}\label{singlediv}
c_1>6\pi\frac{D_1}{g_1}.
\end{equation}
Also numerical solutions were found to diverge beyond this value. A further analysis demonstrates that densities satisfying Eq.~(\ref{singlediv}) lead to the relation $S_1>P_1$. This appears unphysical in the case of polar alignment between the particles of a single species. 

In the two-species case, as expected, $\mathbf{P}_1$ and $\mathbf{P}_2$ show polar alignment for $g>0$ and antiparallel alignment for $g<0$. The general expressions for the static solutions of Eqs.~(\ref{c1P1paradiabatic}) and (\ref{c2P2paradiabatic}) are very lengthy, so we do not list them here. Instead, in the following, we confine ourselves to the special symmetric case of identical species that interact with each other, i.e.~$c_1=c_2$, $g_1=g_2$, and $D_1=D_2$. 

Assuming $P_1=P_2$ and $S_1=S_2$ for the degrees of polar and nematic orientational order, respectively, we obtain the trivial solution $P_1=P_2=0$, or
\begin{equation}\label{P1eqP2}
P_1^2=P_2^2= 4\pi D_1\frac{(g_1+|g|)c_1-2\pi D_1}{(g_1+|g|)^2c_1^2}.
\end{equation}
The latter implies
\begin{equation}
S_1=S_2= 1-\frac{2\pi D_1}{(g_1+|g|)c_1}.
\end{equation}
As we can see from Eq.~(\ref{P1eqP2}), the nontrivial solution exists for densities
\begin{equation}\label{critcP1eqP2}
c_1=c_2> \frac{2\pi D_1}{g_1+|g|}. 
\end{equation}

We performed a linear stability analysis in $P_1=P_2$ and $S_1=S_2$, in the angular orientations of $\mathbf{P}_1$ and $\mathbf{P}_2$, as well as in the angular orientations of the principal axes of $\mathbf{Q}_1$ and $\mathbf{Q}_2$. This linear stability analysis confirmed the critical density values of Eq.~(\ref{critcP1eqP2}). They are the analog to the critical particle densities for the single-species system as they follow from Eq.~(\ref{critrhosep}). Below this density value, the system disorders to the state $P_1=P_2=0$. On the other hand, we found that the solution becomes linearly unstable for densities
\begin{equation}
c_1=c_2> 6\pi\frac{D_1}{g_1+|g|}.
\end{equation}
This is the analog to expression (\ref{singlediv}) of the single-species case, where the macroscopic equations are not stable any more. $g_1$ is replaced by the stronger coupling $g_1+|g|$.

\subsection{Perpendicular alignment}

The situation can be manifestly different in the case of perpendicular alignment, $a=-2$, of Eq.~(\ref{Ugperp}). Following the same procedure as in the previous section, we now find
\begin{eqnarray}\label{c1P1perp}
\partial_t(c_1\mathbf{P}_1) & \approx & 
{}-2u_1\left[\nabla\cdot(c_1\mathbf{Q}_1)+\frac{1}{2}\nabla c_1\right]
-D_1c_1\mathbf{P}_1 \nonumber\\
&& {}+\frac{g_1}{\pi}c_1^2\left[\frac{1}{2}\mathbf{P}_1-\mathbf{P}_1\cdot\mathbf{Q}_1\right] \nonumber\\[.1cm]
&& {}-\frac{g}{\pi}c_1c_2\mathbf{P}_1\cdot\mathbf{Q}_2
\end{eqnarray}
and 
\begin{eqnarray}\label{c2P2perp}
\partial_t(c_2\mathbf{P}_2) & \approx & 
_1\leftrightarrow\hspace{.4pt}_2
\end{eqnarray}
for the polar alignment vectors as well as 
\begin{eqnarray}\label{c1Q1perp}
\partial_t(c_1\mathbf{Q}_1) & \approx & 
\frac{u_1}{2}\left[\nabla\cdot(c_1\mathbf{P}_1)\right]\mathbf{I} \nonumber\\
&& {}-\frac{u_1}{2}\left[\nabla (c_1\mathbf{P}_1)+\left\{\nabla(c_1\mathbf{P}_1)\right\}^T\right] \nonumber\\[.1cm]
&& {}+\frac{g_1}{2\pi}c_1^2\left[2\mathbf{P}_1\mathbf{P}_1-\mathbf{P}_1\!\!^2\,\mathbf{I}\right] \nonumber\\
&& {}-4D_1c_1\mathbf{Q}_1-\frac{g}{\pi}c_1c_1\mathbf{Q}_2
\end{eqnarray}
and 
\begin{eqnarray}\label{c2Q2perp}
\partial_t(c_2\mathbf{Q}_2) & \approx & 
_1\leftrightarrow\hspace{.4pt}_2
\end{eqnarray}
for the nematic order parameters. 

In contrast to the previous case of polar and antiparallel alignment, the nematic order parameters $\mathbf{Q}_1$ and $\mathbf{Q}_2$ are now explicitly coupled in Eqs.~(\ref{c1Q1perp}) and (\ref{c2Q2perp}). This is a consequence of the different angular interaction potential (\ref{Ugperp}) between the two species. It is of second order in the angular momenta. 

Since the angular orientation of one of the polar vector order parameters is arbitrary, we can reduce the number of independent variables to seven. Still, however, a general systematic stability analysis is out of reach and we discuss special stationary solutions in the following. 

Proceeding in the same way as in the previous subsection, the stationary spatially homogeneous values of $\mathbf{Q}_1$ and $\mathbf{Q}_2$ now become 
\begin{eqnarray}
\mathbf{Q}_{1,st} & \approx & \frac{\pi}{16\pi^2D_1D_2-g^2c_1c_2} \nonumber\\
&& {}\times\bigg\{2g_1D_2c_1\left[2\mathbf{P}_1\mathbf{P}_1-\mathbf{P}_1\!\!^2\,\mathbf{I}\right] \nonumber\\
&& \qquad{}-\frac{gg_2}{2\pi}c_2^2\left[2\mathbf{P}_2\mathbf{P}_2-\mathbf{P}_2\!\!^2\,\mathbf{I}\right]\bigg\}, 
\end{eqnarray}
and
\begin{eqnarray}
\mathbf{Q}_{2,st} & \approx & _1\leftrightarrow\hspace{.4pt}_2.
\end{eqnarray}
We can see that the expression diverges for $c_1c_2=16\pi^2D_1D_2/g^2$. This indicates that another solution sets in at such density values. Indeed, from Eq.~(\ref{mod2instab}), our linear stability analysis shows us that the second mode of orientational order becomes unstable at these densities. It corresponds to purely nematic order. For the moment, we confine ourselves to the density regime below this divergence, 
\begin{equation}\label{c1c2small}
c_1c_2 < \frac{16\pi^2D_1D_2}{g^2}.
\end{equation}

First, we assume that one of the two densities is so small that only the other species moves collectively. We choose $\mathbf{P}_1\neq\mathbf{0}=\mathbf{P}_2$. For the magnitude of polar orientational order we find 
\begin{equation}
P_1^2=\frac{(g_1c_1-2\pi D_1)(16\pi^2D_1D_2-g^2c_1c_2)}{g_1c_1^2(4\pi D_2g_1-g^2c_2)}.
\end{equation}
If all terms on the right-hand side are positive, the solution exists. This is the case for $c_1$ above the critical single-species density and according to Ineq.~(\ref{c1c2small}) if the denominator is positive. The latter is true if
\begin{equation}\label{conditiononeP0}
c_2<\frac{4\pi D_2g_1}{g^2}.
\end{equation}
Otherwise $P_1$ diverges and then becomes imaginary. 

The corresponding degree of nematic orientational order reads
\begin{equation}
S_1 = \frac{\pi D_2(g_1c_1-2\pi D_1)}{c_1(4\pi g_1D_2-g^2c_2)}, 
\end{equation}
with the principal axis of $\mathbf{Q}_1$ oriented parallel to $\mathbf{P}_1$. Interestingly, the coupling induces nonzero nematic orientational order of species $2$ that is of strength
\begin{equation}
S_2 = \frac{g(g_1c_1-2\pi D_1)}{4(4\pi g_1D_2-g^2c_2)}.
\end{equation}
The principal axis of $\mathbf{Q}_2$ is oriented perpendicular to the one of $\mathbf{Q}_1$ and to $\mathbf{P}_1$. 

Looking for solutions of non-vanishing polar order for both species, $\mathbf{P}_1\neq\mathbf{0}\neq\mathbf{P}_2$, we find that the relative angle between the two vectors is $\pm\frac{\pi}{2}$. The magnitudes of the vectors are given by 
\begin{eqnarray}
P_1^2 & = & \frac{1}{g_1c_1^2(g_1g_2-g^2)} \nonumber\\[.1cm]
&& {}\times\bigg\{ (4\pi g_2D_1-g^2c_1)(g_1c_1-2\pi D_1) \nonumber\\
&& \qquad{}+g(4\pi D_1-g_1c_1)(g_2c_2-2\pi D_2) \bigg\}
\end{eqnarray}
and
\begin{eqnarray}
P_2^2 & = & _1\leftrightarrow\hspace{.4pt}_2.
\end{eqnarray}
This solution trivially exists, if the signs of all the terms in brackets are positive. Above the single-species thresholds $c_j>2\pi D_j/g_j$ ($j=1,2$), this is achieved if simultaneously $c_1<4\pi g_2D_1/g^2$, $c_2<4\pi g_1D_2/g^2$ [same condition as Ineq.~(\ref{conditiononeP0})], 
\begin{equation}
c_j<\frac{4\pi D_j}{g_j}, \qquad j=1,2,
\end{equation}
and
\begin{equation}
g^2<g_1g_2.
\end{equation}
Indeed we numerically found divergence of the set of Eqs.~(\ref{c1P1perp})--(\ref{c2Q2perp}) when the latter condition was violated. The corresponding nematic order parameters read
\begin{equation}
S_1 = \frac{g_2(g_1c_1-2\pi D_1)+g(g_2c_2-2\pi D_2)}{c_1(g_1g_2-g^2)}
\end{equation}
and
\begin{eqnarray}
S_2 & = & _1\leftrightarrow\hspace{.4pt}_2, 
\end{eqnarray}
where we assumed that the principal axis of $\mathbf{Q}_j$ is parallel to $\mathbf{P}_j$ to choose the sign of $S_j$ ($j=1,2$). 

At the threshold indicated by Eq.~(\ref{c1c2small}), namely $c_1c_2=16\pi^2D_1D_2/g^2$, the spatially homogeneous part of equations (\ref{c1P1perp})--(\ref{c2Q2perp}) has the stationary solution $\mathbf{P}_1=\mathbf{0}=\mathbf{P}_2$ as well as arbitrary $\mathbf{Q}_1$ and $\mathbf{Q}_2$. In order to make a statement about the solution above the threshold, nonlinear terms in $\mathbf{Q}_1$ and $\mathbf{Q}_2$ are necessary. We therefore rederived the set of equations (\ref{c1P1perp})--(\ref{c2Q2perp}) including orientational moments up to fourth order. The third and the fourth modes were used to close the equations. It turned out that in the case that is relevant here, $\mathbf{P}_1=\mathbf{0}=\mathbf{P}_2$ namely, Eq.~(\ref{c1Q1perp}) is supplemented by an expression
\begin{equation}
{}-\frac{g^2}{4\pi^2 D_1}c_1^2c_2^2(\mathbf{Q}_2:\mathbf{Q}_2)\mathbf{Q}_1
\end{equation}
and Eq.~(\ref{c2Q2perp}) accordingly by a corresponding expression of $_1\leftrightarrow\hspace{.4pt}_2$. From this it can be shown that a stationary spatially homogeneous solution of truly nematic order $\mathbf{P}_1=\mathbf{0}=\mathbf{P}_2$ and $\mathbf{Q}_1\neq\mathbf{0}\neq\mathbf{Q}_2$ exists above the threshold given by Eq.~(\ref{c1c2small}) as expected.

\section{Conclusions}

In this paper, we have studied the case of binary mixtures of self-propelled particles. We started from a minimal model in which the magnitude of the particle velocities is kept constant (see, e.g., ref.~\cite{bertin2006boltzmann} for the single-particle case). The orientational order between the particles is adjusted through local pairwise interactions. Polar orientational order is preferred for particles of the same species. For particles of different species, we have investigated the cases of preferred polar, nematic, and perpendicular alignment interactions. 

We started from the Langevin equations in the particle picture. From these, we derived mean field continuum equations of the Fokker-Planck type. The onset of collective motion and the nature of corresponding solutions in the binary self-propelled particle mixtures were studied through a linear stability analysis and numerical investigations of the particle and continuum equations. Furthermore, we derived macroscopic continuum equations and analyzed corresponding stability ranges. 

It turned out that the interaction between the two species can reduce the threshold densities for the onset of collective motion to values below the ones for the single-species case. If one of the two species has a density above this threshold, it moves collectively and can induce collective motion also in the other species even if the latter has a density below the single-species threshold. In the case that both densities are below the single-species threshold, interaction between particles of different species can nevertheless induce collective motion within each species for all three alignment rules investigated. 

Above, but close to the onset of collective motion, spatial heterogeneities in the form of stripe-like flocks emerge in the density profiles. In the most interesting case, one of the species densities is above the single-species threshold and the other is below. Then the first species develops collective motion and spatial heterogeneities. Where its density is high, it can induce collective motion in the second species. Depending on the alignment rules between different species, this can lead to identical or inverted density maps for the two species. Increasing the coupling strength between the two species dissolves the spatial inhomogeneities. 

For the case of preferred perpendicular alignment, we find a competition between polar and truly nematic order as a function of the strength of orientational coupling between the two species. This competition also influences the development of the spatial inhomogeneities. 

When we are looking for the connection of our study to the experimental investigation of real systems, we have to keep in mind that our results were obtained for two spatial dimensions. We should therefore confine ourselves to at least quasi two dimensional systems. Candidates for the latter are thin films of motile bacteria colonies (monolayers in the ideal case) at air-water surfaces or on substrates. For such cultures formed by {\it Bacillus subtilis} it has been shown that only a fraction of the cells is motile, the other cells are non-motile \cite{veening2010gene,kearns2005cell}. The size of this fraction is controlled by the genetic location of the gene responsible for the production of a certain protein \cite{veening2010gene}. In our model system, this situation corresponds to the limiting case in which both species are formed by the same bacterium, one of the species featuring zero motility, the other propelling with non-zero velocity. 

Natural systems in which the latter situation is often observed are bacterial biofilms. These are communities of microorganisms attached to surfaces. Also biofilms of {\it Bacillus subtilis} were demonstrated to feature cellular differentiation so that only part of the cells are motile \cite{veening2010gene,vlamakis2008control}. During the biofilm development, a fraction of the initially motile cells starts to take over a different task and forms non-motile sub-communities \cite{vlamakis2008control}. Although biofilms are typically extended on a surface, their thickness cannot be neglected and can even feature a spatial organization in different layers \cite{vlamakis2008control}. However, very thin films can be produced for example by letting a biofilm of motile bacteria grow in upstream direction in flow cells \cite{houry2010involvement}. 

Despite their abundance in nature and their much higher clinical relevance, multi-species biofilms have only recently been moved into the focus of investigation \cite{elias2012multi}. In this case, different species within the biofilm can interact via quorum sensing and metabolic cooperation or competition. Depending on synergistic or antagonistic interactions between motile bacterial species, different rules of alignment may result when their swarming behavior is investigated. One step into this direction was performed by a study on a community of two different species of motile bacteria, in which the authors also focus on the role of motility on species interactions within a biofilm \cite{an2006quorum}. 

An aspect that has not been addressed in our work by the mean field approach is the nature and role of density fluctuations. For the case of a single particle species it has been demonstrated that large fluctuations in the density can occur \cite{toner1998flocks,chate2008modeling,chate2008collective,peruani2011polar}. These have been termed giant number fluctuations in the context of nematic particle interactions \cite{ramaswamy2003active}. Questions that arise are, for example, how the nature of these fluctuations changes in the case of binary mixtures of self-propelled particles, whether and how the two particle species interact through density fluctuations, and how the situation changes when different alignment rules apply. 

Another issue concerns the macroscopic equations derived in the last section. As discussed, these are stable only for small and moderate particle densities. On the one hand, it will be interesting to find closure relations that stabilize these equations also for higher densities, even if the deviations from the initial equations increase. On the other hand, the non-equilibrium generalization of other transformations from particle to field descriptions that are suitable in the high-density limit is a compelling task for the future. 

In conclusion, here, as a first step, we have shown numerical results for the simplest case where the particles of different species feature the same behavior. That is, single particles of different species propel with the same velocity, show the same orientational diffusion, and follow the same orientational ordering rules. Rich behavior is to be expected when these confining conditions are weakened. Investigations for particle species of different velocities and different magnitudes of orientational diffusion are currently underway.

\acknowledgments

The author acknowledges stimulating discussions with Harald Pleiner, Thorsten St\"uhn, and Burkhard D\"unweg. He thanks Harald Pleiner and Kurt Kremer for a stay in the Theory Group of the Max Planck Institute for Polymer Research in Mainz, where this work was performed.

\end{document}